\def\isfullversion{1}
\def\isfeateval{1}
\def\istargetid{1}

\newif\ifanonymous
\newif\iffullversion
\newif\iffeateval
\newif\iftargetid

\ifdefined\ismodel
\modeltrue
\fi

\ifdefined\isfullversion
\fullversiontrue
\fi

\ifdefined\isfeateval
\featevaltrue
\fi

\ifdefined\istargetid
\targetidtrue
\fi

\ifdefined\isanonymous
\anonymoustrue
\fi

\documentclass[10pt,conference,compsocconf,letterpaper]{IEEEtran}

\usepackage{times}
\usepackage{graphicx}
\usepackage{epstopdf}
\usepackage{enumitem}
\usepackage{xcolor}
\usepackage{color}
\usepackage{booktabs}
\usepackage{multirow}
\usepackage{subcaption}
\usepackage[ampersand]{easylist}
\usepackage{amsmath}

\definecolor{green}{rgb}{0.55, 0.71, 0.0}
\definecolor{red}{rgb}{0.8, 0.25, 0.33}

\iffullversion
\makeatletter
\def\@copyrightspace{\relax}
\makeatother
\fi

\ifCLASSOPTIONcompsoc
  \usepackage[nocompress]{cite}
\else
  \usepackage{cite}
\fi

\iffullversion
\IEEEtriggeratref{33}
\else
\IEEEtriggeratref{21}
\fi

\begin{document}

\title{Know Your Phish: Novel Techniques for Detecting Phishing
             Sites and their Targets}



%
\author{\IEEEauthorblockN{Samuel Marchal\IEEEauthorrefmark{1},
Kalle Saari\IEEEauthorrefmark{1},
Nidhi Singh\IEEEauthorrefmark{2} and
N. Asokan\IEEEauthorrefmark{1}\IEEEauthorrefmark{3}}
\IEEEauthorblockA{\IEEEauthorrefmark{1}Aalto University}
\IEEEauthorblockA{\IEEEauthorrefmark{2}Intel Security}
\IEEEauthorblockA{\IEEEauthorrefmark{3}University of Helsinki\\
Email: \{samuel.marchal,kalle.saari\}@aalto.fi; nidhi.singh@intel.com; asokan@acm.org}
}


\maketitle
\IEEEpeerreviewmaketitle

\newcommand{\eat}[1]{}

\begin{abstract}
Phishing is a major problem on the Web. Despite the significant attention it has received over the years, there has been no definitive solution. While the state-of-the-art solutions have reasonably good performance, they require a large amount of training data and are not adept at detecting phishing attacks against new targets.

In this paper, we begin with two core observations: (a) although phishers try to make a phishing webpage look similar to its target, they do not have unlimited freedom in structuring the phishing webpage; and (b) a webpage can be characterized by a small set of key terms; how these key terms are used in different parts of a webpage is different in the case of legitimate and phishing webpages. Based on these observations, we develop a phishing detection system with several notable properties: it requires very little training data, scales well to much larger test data, is language-independent, fast, resilient to adaptive attacks and implemented entirely on client-side. In addition, we developed a target identification component that can identify the target website that a phishing webpage is attempting to mimic. The target detection component is faster than previously reported systems and can help minimize false positives in our phishing detection system.

\end{abstract}

\section{Introduction}

Phishing webpages (``phishs'') lure unsuspecting web surfers into revealing their credentials. As a major security concern on the web, phishing has attracted the attention of many researchers and practitioners. There is a wealth of literature, tools and techniques for helping web surfers to detect and avoid phishing webpages. Nevertheless, phishing detection remains an arms race with no definitive solution. State-of-the-art large scale real-time phishing detection techniques~\cite{whittaker:2010:large} are capable of identifying phishing webpages with high accuracy (\textgreater 99\%) while achieving very low rates of misclassifying legitimate webpages (\textless 0.1\%).
However, many of these techniques, which use machine learning, rely on millions of static features, primarily taking the bag-of-words approach. This implies two major weaknesses: (a)  they need a huge amount of labeled data to train their classification models; and (b) they are language- and brand-dependent and not very effective at identifying new phishing webpages targeting brands that were not already observed in previous attacks. Commercial providers of phishing detection solutions struggle with obtaining and maintaining labeled training data. From the deployability perspective, solutions that require minimal training data are thus very attractive.

In this paper, we introduce a new approach that avoids these drawbacks. Our goal is to identify whether a given webpage is a phish, and, if it is, identify the \emph{target} it is trying to mimic.
Our approach is based on two core conjectures:
\begin{itemize}
\item\textbf{Modeling phisher limitations}: To increase their chances of success, phishers try to make their phish mimic its target closely and obscure any signal that might tip off the victim. However, in crafting the structure of the phishing webpage, phishers are restricted in two significant ways. First, external hyperlinks in the phishing webpage, especially those pointing to the target, are to domains \emph{outside the control} of phishers. Second, while phishers can freely change most parts of the phishing page, the latter part of its domain name is \emph{constrained} as they are limited to domains that the phishers control. We conjecture that by modeling these limitations in our phishing detection classifier, we can improve its effectiveness. 
\item\textbf{Measuring consistency in term usage}: A webpage can be represented by a collection of key terms that occur in multiple parts of the page such as its body text, title, domain name, other parts of the URL etc. We conjecture that the way in which these terms are used in different parts of the page will be different in legitimate and phishing webpages.
\end{itemize}

Based on these conjectures, we develop and evaluate a phishing detection system. We use comparatively few (212) but relevant features. This allows our system, even with very little labeled training data, to have high accuracy and low rate of mislabeling legitimate websites. By modeling inherent phisher limitations in our feature set, the system is resilient to adaptive attackers who dynamically change a phish to circumvent detection.  
Our basic phishing detector component
(Section~\ref{sec:classification}) does not require online access to
centralized information and is fast. Therefore, it is highly suited
for a privacy-friendly client-side implementation. Our target brand
identification component (Section~\ref{sec:target}) uses a simple
technique to extract a set of \emph{keyterms} characterizing a webpage
and, in case it is a phish, uses the keyterms set to identify its
target. Both components eschew the bag-of-words approach and are thus
not limited to specific languages or targeted brands.
\newpage
We claim the following contributions:

\begin{itemize}
	\item a new set of features to detect phishing webpages (Section~\ref{subsec:feat_comp}) and a classifier, using these features, with the following properties that distinguish it from previous work:
          \begin{itemize}
          \item it learns a generalized model of phishing and legitimate webpages from \textbf{a small training set} (few thousands).
          \item it is \textbf{language- and brand-independent}.
          \item it is resilient to adaptive attackers.
          \item its features are extracted only from information retrieved by a web browser from the webpage and it does not require online access to centralized information. Hence it \textbf{admits a client-side-only implementation} that offers several advantages including (a) better privacy, (b) real-time protection and (c) resilient to phishing webpages that return different contents to different clients.
          \end{itemize} 
        \item comprehensive evaluation of this system, showing that its accuracy (\textgreater 99\%) and misclassification rate (\textless 0.1\%) are comparable to prior work while using significantly smaller training data. (Section~\ref{subsec:exp_classif}) 
	\item a fast target identification technique (Section~\ref{sec:target}) for phishing webpages with accuracy (90-97\%) comparable to previously reported techniques. It can also be used to remove false positives from the basic phishing detection component described above.  (Section~\ref{subsec:target_eval})
\end{itemize}
\iffullversion
This research report is an extended version of an ICDCS 2016 paper \cite{marchal:2016:know}. A proof of concept of this technique has been implemented as a phishing prevention browser add-on \cite{armano:2016:real}  
\fi

\section{Background}
\label{sec:background}

\subsection{Phishing}
\label{subsec:phishing}
Phishing refers to the class of attacks where a victim is lured to a fake webpage masquerading as a target website and is deceived into disclosing personal data or credentials. Phishing campaigns are typically conducted using spam emails to drive users to fake websites \cite{apwg:2015:2}.
Impersonation techniques range from technical subterfuges (email spoofing, DNS spoofing, etc.) to social engineering. The former is used by technically skilled phishers while unskilled phishers resort to the latter~\cite{hardy:2014:targeted}. 
Phishing webpages mimic the look and feel of their target websites~\cite{Xiang:2009:hybrid}. In order to make the phishing webpages believable, phishers may 
embed some content (HTML code, images, etc.) taken directly from the target website and use relatively little content that they themselves host \cite{pan:2006:anomaly}. This includes outgoing links pointing to the target website.
They also use keywords referring to the target in different elements of the phishing webpage (title, text, images, links)~\cite{apwg:2015:2,le:2011:phishdef,marchal:2014:phishstorm,pan:2006:anomaly}. 
In this paper, our focus is on detection of phishing webpages created by an attacker and hosted on his own web server or on someone else's compromised web server.


\subsection{URL Structure}
\label{subsec:url_structure}

\begin{figure}[htpb]
        \centering
        \includegraphics[width=0.8\linewidth]{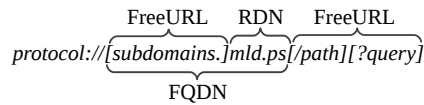}
        \caption{Structure of a URL}\label{fig:url}
\end{figure}

\iffeateval
Webpages are addressed by a uniform resource locator (URL).
Fig.~\ref{fig:url} shows relevant parts in the structure of a typical URL. It begins with the \textit{protocol} used to access the page. The fully qualified domain name (\textit{FQDN)} identifies the server hosting the webpage. It consists of a registered domain name (\textit{RDN}) and prefix which we refer to as \textit{subdomains}. A phisher has full control over the \textit{subdomains} portion and can set it to any value. The \textit{RDN} portion is constrained since it has to be registered with a domain name registrar. \textit{RDN} itself consists of two parts: a \textit{public suffix}\footnote{Public Suffix List (https://publicsuffix.org/)} (\textit{ps}) preceded by a \textit{main level domain} (\textit{mld}). The URL may also have a \textit{path} and \textit{query} components which, too, can be changed by the phisher at will. We use the term \textit{FreeURL} to refer to those parts of the URL that are fully controllable by the phisher.
\else
Webpages are addressed by a uniform resource locator (URL).
Fig.~\ref{fig:url} shows relevant parts in the structure of a typical URL. It begins with the \textit{protocol} used to access the page. The fully qualified domain name (\textit{FQDN)} identifies the server hosting the webpage. It consists of a registered domain name (\textit{RDN}) and prefix which we refer to as \textit{subdomains}. A phisher has full control over the \textit{subdomains} portion and can set it to any value. The \textit{RDN} portion is constrained since it has to be registered with a domain name registrar. \textit{RDN} itself consists of two parts: a \textit{public suffix} (\textit{ps}) preceded by a \textit{main level domain} (\textit{mld}). The URL may also have a \textit{path} and \textit{query} components which, too, can be changed by the phisher at will. We use the term \textit{FreeURL} to refer to those parts of the URL that are fully controllable by the phisher.
\fi

Consider an example URL:\\
\textit{https://www.amazon.co.uk/ap/signin?\_encoding=UTF8}\\
We can identify the following components:
\begin{itemize}
\item \emph{protocol} = \textit{https}
\item \emph{FQDN} = \textit{www.amazon.co.uk}
\item \emph{RDN} = \textit{amazon.co.uk}
\item \emph{mld} = \textit{amazon}
\item \emph{FreeURL} = \{\textit{www}, \textit{/ap/signin?\_encoding=UTF8}\}
\end{itemize}

\subsection{Data Sources}
\label{subsec:data_source}

From analyzing phishing webpages, we identify the following data sources, available to a web browser when it loads a webpage, that can be useful in detecting phishing webpages:

\begin{itemize}
	\item \emph{Starting URL}: the URL given to the user to access the website. 
It can be distributed in emails, instant messages, websites, documents, etc.
	\item \emph{Landing URL}: the final URL pointing to the actual content presented to the user in his web browser. This is the URL present in the browser address bar when the page is completely loaded.
	\item \emph{Redirection chain}: the set of URLs crossed to go from the starting URL to the landing URL (including both).
	\item \emph{Logged links}: the set of URLs logged by the browser while loading the page. They point to sources from which embedded content (code, images, etc.) in the webpage are loaded.
	\item \emph{HTML}: the HTML source code of the webpage and IFrames included in the page. We consider four elements extracted from this source code:
	\begin{itemize}
		\item \emph{Text}: text contained between \textit{\textless body\textgreater} HTML tags (actually rendered on user's display).
		\item \emph{Title}: text contained between \textit{\textless title\textgreater} HTML tags (appears in the browser tab title).
		\item \emph{HREF links}: the set of URLs representing outgoing links in the webpage.
		\item \emph{Copyright}: the copyright notice, if any, in Text.
	\end{itemize}
	\item \emph{Screenshot}: an image capture of the loaded webpage. 
\end{itemize}
\section{Design Overview}
\label{sec:design}

\iffeateval
Before describing the detailed design (in Sections~\ref{sec:classification} and \ref{sec:target}), we start with an overview.
\fi

\subsection{Modeling Phisher Limitations}
\label{subsec:phisher_limitation}

In Section~\ref{subsec:url_structure}, we saw that even on systems they control, phishers are constrained from freely constructing URLs to pages they host. Similarly, in Section~\ref{subsec:phishing}, we saw that in order to maximize the believability of their phishing sites, phishers include content from URLs outside their control. 
\iffullversion
We conjecture that by taking these constraints and level of control into account in selecting and grouping features for our classification, we can improve classification performance.
\fi
Thus, we divide the data sources from Section \ref{subsec:data_source} into subcategories according to the level of \textit{control} phishers may have on them and the \textit{constraints} on phishers.

\noindent\textbf{Control}: URLs from \textit{logged links} and \textit{HREF links} are subdivided into \textit{internal} and \textit{external} according to their \textit{RDN}.
The set of \textit{RDN}s extracted from URLs involved in the redirection chain are assumed to be under the control of the webpage owner. Any URLs that include these \textit{RDN}s are marked \textit{internal}. Other \textit{RDN}s are assumed to be possibly outside the control of the webpage owner. URLs containing such \textit{RDN}s are marked \textit{external}.

\noindent\textbf{Constraints}:
Within a URL, we distinguish between \textit{RDN}, which cannot be freely defined by the webpage owner, and (\textit{FreeURL}), which can be.

\subsection{Extracting Term Distributions}
\label{subsec:labels_set}


The primary technique of a phisher is essentially social engineering: fooling a victim into believing that the phishing webpage is the target~\cite{hardy:2014:targeted}. Thus, it is plausible that lexical analysis of the data sources will help in identifying phishing webpages: we conjecture that legitimate webpages and phishing webpages differ in the way terms are used in \textit{different locations} in those pages.
To incorporate measurements of such term usage consistency, we first define what ``terms'' are and how they are extracted from a webpage.
Let $A$ be the set of the 26 lowercase English letters:
$A = \left\lbrace a, b, c, ..., x, y, z \right\rbrace$. We extract terms from a data source as follows:
\begin{itemize}
 \item canonicalize letter characters by mapping upper case characters, accented characters and special characters to a matching letter in $A$; e.g., \textit{ \{ B, $\beta$, \`b, \^b \} }$\rightarrow b$.
 \item split the input into substrings whenever a character  $c \notin A$ is encountered.
 \item throw away any substring whose length is less than 3.
\end{itemize}
Let  $T = A^n \vert n \geq 3$ be the set of all possible terms. Suppose $T_S = \{t_{i \in \left\lbrace 1;m \right\rbrace} \in T\}$ was extracted from a data source $S$ and $t_i$ occurs with probability $p_i$.
The set of $m$ pairs $(t_i, p_i ) \in T \times \left] 0,1 \right], i \in \left\lbrace 1;m \right\rbrace $ represents the \emph{term distribution} $D_{S}$ of \emph{S}.

\begin{table}[tbh]
\caption{Term distributions} 
\centering
\begin{tabular}{l l}

\textbf{Distribution} & \textbf{Data source} \\ \hline
$D_{text}$ & Text \\
$D_{title}$ & Title \\
$D_{copyright}$ & Copyright notice  \\
$D_{image}$ & Webpage screenshot \\
$D_{start}$ & Starting URL  -- \textit{FreeURL} \\
$D_{land}$ & Landing URL -- \textit{FreeURL} \\
$D_{intlog}$ & Internal logged links -- \textit{FreeURL} \\
$D_{intlink}$ & Internal HREF links -- \textit{FreeURL} \\
$D_{startrdn}$ & Starting URL -- \textit{RDN} \\
$D_{landrdn}$ & Landing URL -- \textit{RDN} \\
$D_{intrdn}$ & Internal links (HREF and logged) -- \textit{RDN} \\
$D_{extrdn}$ & External logged links -- \textit{RDN} \\
$D_{extlog}$ & External logged links -- \textit{FreeURL} \\
$D_{extlink}$ & External HREF links -- \textit{FreeURL} \\

\end{tabular}
\label{tab:label_sets}
\end{table}


Table \ref{tab:label_sets} defines the term distributions we consider.
The external sources $extrdn$, $extlog$, $extlink$ are those assumed to be outside the control of the webpage owner. \textit{RDN} data sources $startrdn$, $landrdn$, $intrdn$ are constrained by DNS registration. The rest is controlled by the webpage owner without constraints.
\iffullversion
Table~\ref{tab:set_constraints} summarizes the level of control and the constraints a webpage owner has on the data sources of a webpage.
\fi
The $image$ data source is composed of terms extracted by optical character recognition (OCR) from the screenshot of a rendered webpage. 
\iffeateval
It is used to address the case of image-based webpages from which no text content can be extracted.
\fi



\iffullversion
 \begin{table}[tbh]
 \centering
 \caption{Data sources control and constraints}
 \label{tab:set_constraints}
 \begin{tabular}{l l l}
 								& \textbf{Controlled} & \textbf{Uncontrolled} \\ \hline
 \multirow{8}{*}{\textbf{Unconstrained}} & $text$ & $extlog$ \\
 								& $title$ & $extlink$ \\
 								& $copyright$  & \\
 								& $image$ & \\ 
 								& $start$ & \\
 								& $land$ & \\
 								& $intlog$ & \\
 								& $intlink$ & \\ \hline								
 								& $startrdn$ & $extrdn$ \\
 \textbf{Constrained}	& $landrdn$ & \\
 								& $intrdn$ & \\								
 
 \end{tabular}
 \end{table}
 \fi

\subsection{Architecture}
Our overall design consists of a phishing webpage detection system (Section~\ref{sec:classification}) and a target identification system (Section~\ref{sec:target}). The phishing detection system is a classifier that identifies phishing webpages based on a set of newly introduced features. 
The target identification system identifies if a given webpage is a phish by finding its target. 
Both systems can be used in a pipeline: the phishing detection system tentatively identifies a potential phish, which can be fed to the target identification system to infer the purported target.
\iffeateval
If a target is not found, then it can be deemed a false positive.
\fi
\section{Phishing Detection System}
\label{sec:classification}

\iffeateval
We now describe how and why we select the set of features we use in our phishing detection classifier.
\fi
  
\subsection{Feature Set Requirements}
\label{subsec:requirements}

We consider some facts of phishing detection in order to deduce requirements that a feature set must have:

\noindent\textbf{Generalizability}: Accumulating ground truth phishing and legitimate data is challenging. Phishing websites have very short lifetimes~\cite{apwg:2015} and can display different content depending on a browser's user-agent or user's geographic location. Labeled phishing and legitimate resources are often defined by URLs (\textit{e.g.} PhishTank\footnote{PhishTank (https://www.phishtank.com/)}). Assigning correct labels (phish, non-phish) to these URLs is difficult. But even if the initial labeling was done correctly, information on the pages pointed by them can also evolve over time: a legitimate domain name can be hijacked to host phishing content for a while or a phishing domain name can be parked or changed to contain empty content after a short uptime. Therefore, crawling a set of labeled URLs to gather ground truth data often leads to noisy datasets that further require manual checking and cleaning up.
Thus it is desirable to select a feature set that allows a model to be learned from \emph{as small a training set as possible} while remaining applicable to far larger test datasets. 
Using a much larger test set than the training set also allows the detection and avoidance of overfitting \cite{Dietterich:1995:overfitting}.  

\noindent\textbf{Adaptability}: Several automated classification techniques \cite{thomas:2011:design,whittaker:2010:large,zhang:2007:cantina} rely on a static set of features learned from a training set such as the bag-of-words model or ``term frequency-inverse document frequency'' (TF-IDF) \cite{Salton:1983:introduction} computation. Such feature models are language-dependent and vary with training sets. Using such features shows \cite{whittaker:2010:large} that certain terms such as \textit{paypal} are dominant features. Thus the efficacy of such models on phishs that masquerade as previously unknown targets or brands is questionable. 
In addition, phishers can adaptively modify the content of their phish to circumvent detection by such static models, e.g.,  by using words that typically occur in legitimate webpages.
An adaptable feature set must be \emph{independent of learning instances}, preferably defined manually with motivated reasons, and be resilient to adaptive attacks.

\noindent\textbf{Usability}: It is desirable that features are computable on an end user system \emph{without relying on online access to centralized servers} or proprietary data (\textit{e.g.} Google PageRank). This preserves user privacy since the scheme does not require users to disclose their browsing history to an outside entity. 


\noindent\textbf{Computational Efficiency}: Features must be \emph{quickly computable} to allow integration with real time detection systems that do not impact users' web surfing experience.

\subsection{Computing Features}
\label{subsec:feat_comp}

We now introduce 212 features and motivate their selection. 
We intend to capture the constraints and degree of control discussed earlier (Section~\ref{subsec:phisher_limitation}) as well as consistency checking of term usage (Section~\ref{subsec:labels_set}).
We group features into five categories (Table \ref{tab:feature_type}).

\begin{table}[tbh]
\caption{Feature sets} \centering
\begin{tabular}{r r l}

\textbf{Name} & \textbf{Count}  & \textbf{Type}\\ \hline
$f_1$ & 106 & URL \\
$f_2$ &66 & Term usage consistency \\
$f_3$ &22 & Usage of starting and landing \textit{mld} \\
$f_4$ &13 & \textit{RDN} usage \\
$f_5$ &5 & Webpage content \\ \hline
$f_{all}$ & 212 & Entire feature set

\end{tabular}
\label{tab:feature_type}
\end{table}

\noindent\textbf{URL}: 
First we  define nine statistical features related to the lexical composition of URLs (Table \ref{tab:url_features}). Feature $2$ is meant to identify strings in \emph{path} and \emph{query} that look like domain names. Phishing URL and domain name obfuscation techniques \cite{le:2011:phishdef} tend to produce long URLs composed of many terms. This is the rationale for features $3$-$8$. The popularity rank of the domain (feature $9$) is based on a fixed, previously downloaded list of the Alexa top million domain names\footnote{Alexa (http://www.alexa.com/)}. If a domain is not in this list, feature $9$ takes the default value of 1,000,001.

All nine features are extracted from the starting URL (9) and landing URL (9). The mean, median and standard deviation values are computed for features $3$-$9$ on the following sets of URLs: internal logged links, external logged links, internal HREF links and external HREF links ($4*7*3$). Feature $1$ is computed on these sets as a ratio of URLs using \textit{https} over the total count of URLs for each set ($4*1$). Feature $2$ is computed only for the starting and landing URLs.
\iffullversion
(Since URL obfuscation techniques are effective only on URLs that are visible to the user)
\fi
Thus, the complete URL-based feature set ($f_1$) consists of 106 features: $9+9+ 4* (7*3 + 1) = 106$.

\begin{table}[tbh]
\caption{URL features} \centering
\begin{tabular}{l l}

\# & \textbf{Description} \\ \hline
1 & protocol used (http/https) \\
2 & count of dots \textit{`.'} in \textit{FreeURL} \\
3 & count of level domains \\
4 & length of the URL \\
5 & length of the \textit{FQDN} \\
6 & length of the \textit{mld} \\
7 & count of terms in the URL \\
8 & count of terms in the \textit{mld} \\
9 & Alexa ranking of the \textit{RDN}

\end{tabular}
\label{tab:url_features}
\end{table}

\noindent\textbf{Term usage consistency}: The second set of features ($f_2$) captures the consistency of term usage between different types (controlled vs. uncontrolled; constrained vs. unconstrained) of data sources in the page. 
\iffullversion
Since phishers cannot freely define all elements of the webpage they serve to their victims, they obfuscate only some elements by using key terms that will deceive them. 
\fi
Using 12 term distributions (we discard $D_{copyright}$ and $D_{image}$) defined in Section \ref{subsec:labels_set} we define 66 features ($12*11/2$) depicting the similarity of 
pairs of sources by computing pairwise Hellinger Distance between their distributions.
The Hellinger Distance \cite{cam:2000:asymptotics} is a metric used to quantify the dissimilarity between two probabilistic distributions $P$ and $Q$. It is an instance of \textit{f}-divergence that is symmetric and bounded in $\left[ 0,1 \right] $. The value $1$ represents complete dissimilarity ($P \cap Q = \emptyset$) and the value $0$ means that  $P$ and $Q$ are the same probabilistic distribution. 
\iffeateval
The formula of the Hellinger Distance in discrete space is given in Equation \eqref{eq:hellinger}.

\begin{equation}
H^{2}(P,Q) = \frac{1}{2} \sum_{x\in P\cup Q}\left(\sqrt{P(x)} - \sqrt{Q(x)}\right) ^{2}
\label{eq:hellinger}
\end{equation}
\fi

\noindent\textbf{Usage of starting and landing \textit{mld}}: Legitimate websites are likely to register a domain name reflecting the brand or the service they represent. However, phishers often use domain names having no relation with their target \cite{Xiang:2009:hybrid}. Hence, we expect the starting \textit{mld} and/or the landing \textit{mld} to appear in several sources extracted from a legitimate webpage while phishing webpages should not have this characteristic.  
We define 22 features ($f_3$) inferring the usage of the starting and landing \textit{mld} in the text, the title and \textit{FreeURL} of the logged links and HREF links. 
12 binary features are  set to 1 if the starting/landing \textit{mld} appear in $D_{text}$, $D_{title}$, $D_{intlog}$, $D_{extlog}$, $ D_{intlink}$ or $D_{extlink}$ (6*2); 10  features are the sum of probability from terms of $D_{title}$, $D_{intlog}$, $D_{extlog}$, $ D_{intlink}$ and $D_{extlink}$  that are substrings of starting/landing \textit{mld} (5*2). $D_{text}$ is not considered since it is often composed of many short irrelevant terms that match several parts of a \textit{mld}.

\noindent\textbf{\textit{RDN} usage}: We define 13 features ($f_4$) related to \textit{RDN} usage consistency. We compute statistics related to the use of similar and different \textit{RDN}s in starting URL, landing URL, redirection chain, loaded content (logged links) and HREF links. We expect legitimate webpages to use more \textit{internal RDN}s and less redirection than phishing webpages \cite{li:2014:hunting}. 

\noindent\textbf{Webpage content}: Finally, five features ($f_5$) count the number of terms in the text and the title (2), and the number of input fields, images and IFrames (3) in the page. Phishing pages tend to have minimal text to circumvent text-based detection techniques \cite{ramanathan:2013:phishing} and use more images and HTML content loaded from other sources. In addition, since phishing attacks seek to steal user data, phishing webpages often contain several input fields \cite{Xiang:2009:hybrid}.

It is worth noting that while we use terms to compute our feature set, it is not based on any observed language or term usage knowledge. The computation relies solely on the information gathered through a web browser albeit we use a local copy of Alexa ranking list. Hence, it makes the feature set adaptable and usable as well as fast to compute once the data sources are available. 
Since the feature set is small (212) we expect it to have good generalizability.

\subsection{Phishing Detection Model} 

To use our feature set for discriminating phishing from legitimate webpages, we use a supervised machine learning approach. In supervised machine learning, a classification model is learned from observations over a set of data labeled with several classes. The learned model is used to predict the class of unlabeled instances. 
We select Gradient Boosting  \cite{friedman:2002:stochastic} to build the classification model.
\iffeateval
Boosting \cite{freund:1999:ashort} is a fitting procedure that improves the prediction of binary outcomes from weak models using weighted ensembles of base-learners.
It was selected because~\cite{buhlmann:2007:boosting} (a) of its strong ability to select and weight the most relevant features given a set of base learners and (b) boosting algorithms are known to be fairly robust to overfitting, enabling the resulting model to have good generalization capabilities.
\else
It was selected because~\cite{buhlmann:2007:boosting} (a) of its strong ability to select and weight the most relevant features
and (b) boosting algorithms are known to be fairly robust to overfitting, enabling the resulting model to have good generalization capabilities.
\fi

\iffullversion
Gradient Boosting consists in an iterative process where the starting point is an initial estimation of a prediction function that fits poorly the learning data set. During each iteration, the function is improved by fitting a new base-learner (decision tree) to the negative gradient of a pre-specified loss function. Gradient boosting improves model accuracy by selecting variables and weak models at each iteration. At the end of the iterative process, the final prediction function combines the best weak learners and variables as part of the model.
\fi

Gradient Boosting predicts the class of an unknown instance by computing values defined in $\left[ 0,1 \right]$ that gives the confidence of the instance to belong to a given class. In the case of predicting only two classes, the confidence value $v_1$ for one class is equal to $1-v_2$, where $v_2$ is the confidence value for the other class. A \emph{discrimination threshold} predicts, according to the computed confidence values, the class of an instance. By tuning this threshold, we can favor the prediction of one class over the other.
The variation of the discrimination threshold over $\left[ 0,1 \right]$ is used to evaluate the accuracy of a given model by examining how false positive rate varies with true positive rate (ROC) or precision varies with recall.
\section{Target Identification}
\label{sec:target}

The identification of the target of a phishing webpage relies on a set of ``keyterms'' in that webpage related to a brand or service. Rather than leveraging any brand-specific knowledge or text corpus to infer these keyterms \cite{ramesh:2014:efficious,liu:2012:anti}, we introduce a new technique that uses only the information extracted from the webpage. We now discuss in detail how these keyterms are extracted and used.

\subsection{Keyterms Extraction}

A keyterm is one that appears in several data sources (e.g., title, text and landing URL) on a page. We use terms from five data sources introduced in Section \ref{subsec:data_source} that contain user-visible data rendered by the browser:

\begin{itemize}
\item Starting and landing URLs: \\ $T_{start} \cup T_{startrdn} \cup T_{land} \cup T_{landrdn}$
\item Title: $T_{title}$
\item Text: $T_{text}$
\item Copyright: $T_{copyright}$
\item HREF links: $T_{intlink} \cup T_{extlink}$
\end{itemize}

We use three different techniques to identify keyterms. They are used in sequence, depending on the information available in different data sources and the success of each technique (Section~\ref{subsec:identification}).

The first considers the result of pairwise intersection between the five sets of terms as potential keyterms. Each term appearing in at least two data sources is added to a list and ranked in descending order according to a term's overall frequency in the visible parts of the website. 
The top-\textit{N} terms in the ordered list are selected as keyterms. (We use \textit{N=5} in our model as it was proved to be a sufficient number to represent a webpage \cite{zhang:2007:cantina}.) 
These \textit{N} keyterms are called \textit{boosted prominent terms}.

The second technique considers the same data sources but discards the intersection between text and HREF links ($T_{text}  \cap (T_{intlink} \cup T_{extlink})$).
Sometimes text and links of a webpage contain the same terms because the name of the links and the corresponding URL can be the same. This is a common practice in news websites. In this case the intersection terms may be dominated by terms that are irrelevant for target identification and it introduces some noise in the keyterms inference.
The \textit{N} extracted keyterms using this technique are called \textit{prominent terms}.

The last technique applies optical character recognition to the webpage screenshot to produce $T_{image}$. This set of terms is intersected with the five other sets to produce the list of top-\textit{N} keyterms being \textit{OCR prominent terms}. This data source is not used in a first step because OCR is a slow process. Terms extraction from text is much faster. 

\subsection{Identification Process}
\label{subsec:identification}

We now discuss how the extracted keyterms lists (\textit{boosted prominent terms}, \textit{prominent terms} and \textit{OCR prominent terms}) are used to infer the target of a phishing webpage.

\noindent\textbf{Step 1:}
Extract \textit{boosted prominent terms}, and try to ``guess'' the target \textit{FQDN}. The \textit{mld}s from the starting and landing URLs, from the logged links and HREF links are collected. Then, every collected \textit{mld} is checked to figure out if it can be composed based on the keyterms part of \textit{boosted prominent terms} possibly separated by a dash `-' or a string of digits.
\iffullversion
Most commercial URLs are made of a domain name that is their company's name. For instance, the registered domain of Bank of America is \textit{bankofamerica.com}, and the \textit{boosted prominent terms} extracted from the front page of this website include the terms \textit{bank}, \textit{of}, \textit{america}.
\fi
For each guessed \textit{FQDN} (typically 2-3), a search engine query is performed and the returned \textit{RDN}s are stored.
If the \textit{RDN}s of the suspected phishing site (starting and landing URL) appear in the results of the search engine query, we declare this site legitimate and stop the process. Otherwise we go to step 2.
This decision is based on the assumption that a search engine would not return a phishing site as a top hit because (a) a new phishing site (only a few hours old) would not have been indexed by a search engine yet and (b) an old phishing site would have been already detected and ended up in a blacklist.  

\noindent\textbf{Step 2:}
The set of \textit{N prominent terms} is queried against a search engine.
If the suspected \textit{RDN} appears in the set of \textit{RDN}s returned by the search engine, it is declared legitimate and we stop the process.
If some \textit{mld}s resulting form the search engine query appear in a controlled data source of the webpage, we record them and go to step 5. These \textit{mld}s represent the candidate targets. Otherwise, continue to step 3.

\noindent\textbf{Step 3:}
Repeat Step 2 but instead of using \textit{prominent terms}, use \textit{boosted prominent terms}. If the webpage is not confirmed as legitimate and no candidate target is found, go to step 4.

\noindent\textbf{Step 4:}
Repeat Step 2 but instead of using \textit{prominent terms}, use \textit{OCR prominent terms}. If the webpage is not confirmed as legitimate, go to step 5.

\noindent\textbf{Step 5 (target selection):}
For each \textit{mld} candidate target, we count how many times it appears in the data sources of the webpage and rank it in a list according to this criteria.
If a single target is required, we return the most frequent (top-1). If we want to improve the likelihood that the real target is not missed, the top-2 or top-3 most frequent \textit{mld}s can be returned.
\section{Evaluation}

In this section, we present the performance evaluation of the phishing detection system and the target identification method presented in Sections \ref{sec:classification} and \ref{sec:target} respectively. 
\iffullversion
We first describe the experimental setup and datasets used for experiments before evaluating each system individually and presenting the performance of the association of both system. 
\fi

\subsection{Experimental Setup}

Our system is composed of five Python modules:

\noindent\textbf{Webpage scraper} is only required for experiments to gather the information sources defined in Section \ref{subsec:data_source}. It can also be used for offline analysis. The scraper is implemented as a monitored Firefox web browser (Selenium\footnote{Selenium HQ (http://www.seleniumhq.org/)}) that extracts the data sources while visiting a webpage at a given URL. It saves the data in json format and a screenshot of the webpage. 
\iffullversion
Using such setup ensures client-side only implementability.
\fi

\noindent\textbf{Feature extractor} extracts the 212 features (Section \ref{subsec:feat_comp}) from the data sources in the webpage
\iffullversion
gathered by the scraper 
\fi
and builds a feature vector.

\noindent\textbf{Classifier} takes the feature vector and a previously learned classification model as input to predict the class, phishing or legitimate, of a webpage. The implementation of the Gradient Boosting is provided by the Scikit Learn\footnote{Scikit Learn (http://scikit-learn.org/)} Python package. 
\iffeateval
We set the discrimination threshold to $0.7$, which favors the prediction of legitimate webpages ($ \left[0, 0.7 \right[ $) over phishs ($ \left[0.7, 1 \right] $).
\fi

\noindent\textbf{Keyterms extractor} infers the keyterms of a webpage using data gathered by the scraper.

\noindent\textbf{Target identifier} predicts the likelihood of a webpage being a phish. In case of a phish, the modules also identifies its target.

\subsection{Evaluation Datasets}

We obtained URLs from two sources in order to gather ground truth data of phishing and legitimate webpages (Table \ref{tab:dataset}). 
Neither dataset contains personal data. 
\iffullversion
Both datasets are available on request for research use.
\else
We will make both datasets available for research use.
\fi

The phishing URL sets (\textbf{Phish}) were obtained through the community website PhishTank. We conducted three different collection ``campaigns''. The first resulted in \textit{phishTrain} which was used for training the phishing detection classifier. The second, collected at a later point in time, resulted in \textit{phishTest} which was used as the test set. The last, \textit{phishBrand}, was used for evaluating our target identification scheme (Section~\ref{subsec:target_eval}). It consists of 600 phishing webpages for each of which we manually identified the target, resulting in a total of 126 different targets. Each campaign consisted of checking for new entries in PhishTank every hour and scraping the webpages for those URLs.
These are in several languages. The datasets were further manually cleaned to remove any legitimate or unavailable websites and parked domain names. Table \ref{tab:dataset} provides a detailed description of these datasets including the date and the count of elements before and after cleaning.

\begin{table}[tbh]
\caption{Datasets description} \centering
\begin{tabular}{l l l r c}

\textbf{Set} & \textbf{Name} & \textbf{Date (2015)} & \textbf{Initial} & \textbf{Clean}\\ \hline
\textbf{Phish} & \textit{phishTrain} & Jul-23/Aug-3 & 1213 & 1036 \\
& \textit{phishTest} & Sep-13/Sep-24 & 1553 & 1216 \\
& \textit{phishBrand} & Sep-22/Sep-28 & 600 & 600 \\ \hline
\textbf{Leg} & \textit{legTrain} & Jul-15/Jul-22 & 5000 & 4531\\ 
& \textit{English} & Aug-17/Sep-23 & 100,000 & -- \\
& \textit{French} & Sep-28 & 10,000 & -- \\
& \textit{German} & Sep-29 & 10,000 & -- \\
& \textit{Italian} & Sep-30 & 10,000 & -- \\
& \textit{Portuguese} & Oct-1 & 10,000 & -- \\
& \textit{Spanish} & Oct-2 & 10,000 & -- \\

\end{tabular}
\label{tab:dataset}
\end{table}

The legitimate URLs (\textbf{Leg}) were provided by Intel Security\footnote{Intel Security (http://www.intelsecurity.com/)}. We processed them same way as for the phishing URLs.
Intel gave us several datasets. First, an English training set
(\textit{legTrain}) of 5,000 legitimate webpages was cleaned up to remove unavailable websites and dead links. 
Six larger test sets of webpages in different languages (English, French, German, Portuguese, Italian and Spanish) were gathered and did not receive any cleaning treatment. A detailed description of these sets is provided in Table \ref{tab:dataset} as well.
The variety and popularity of the URLs in the test set is reflected in the fact that 65,302 (43.5\%) of the 150,000 test URLs in \textbf{Leg} have \textit{RDN}s ranked in Alexa top 1M.

\iffullversion
\begin{figure*}[th]
        \begin{subfigure}[b]{0.33\textwidth}
                \centering
                \includegraphics[width=\textwidth]{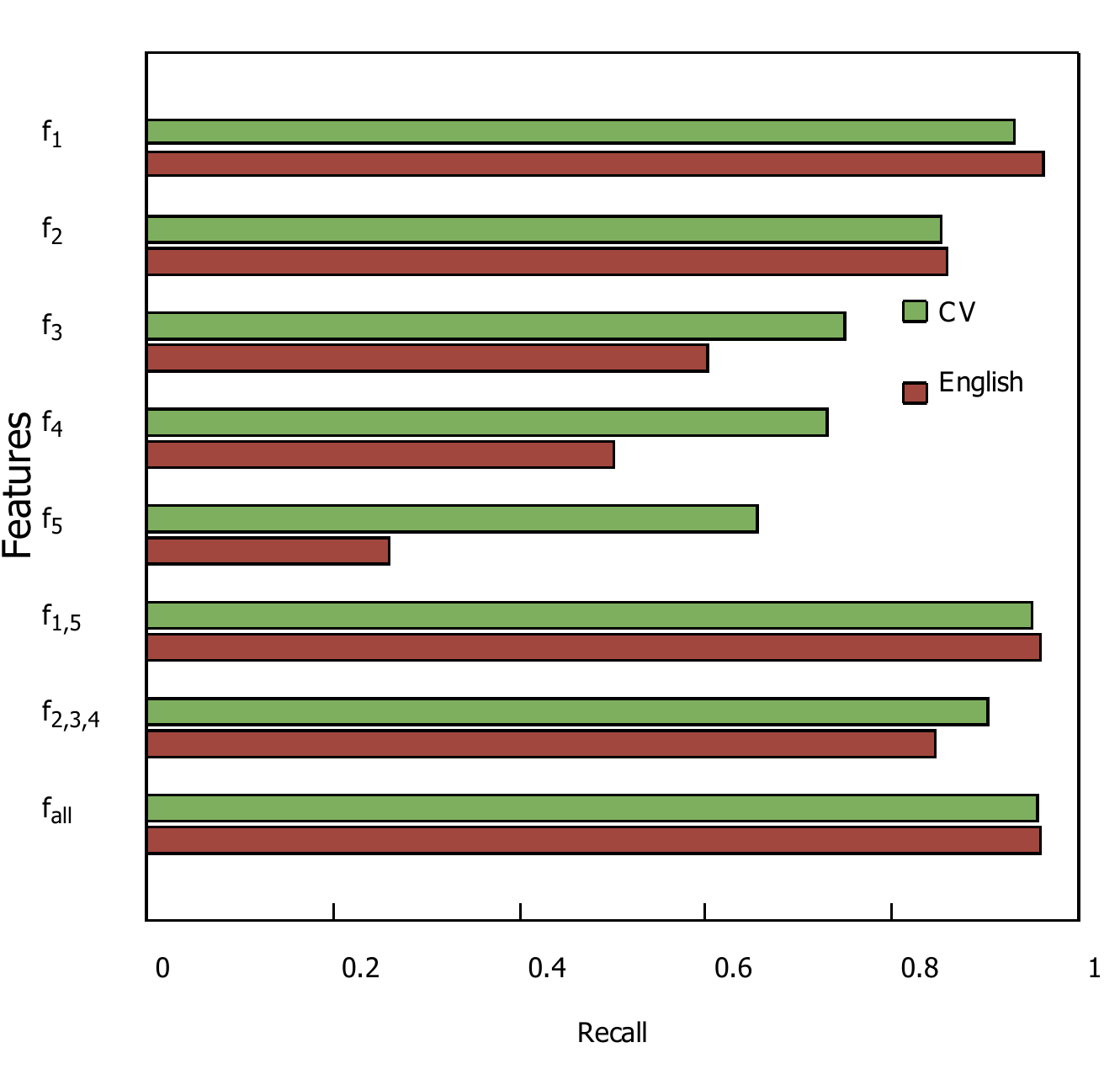}
                \caption{}
                \label{fig:recall}
        \end{subfigure}%
        ~ 
        \begin{subfigure}[b]{0.33\textwidth}
                \centering
                \includegraphics[width=\textwidth]{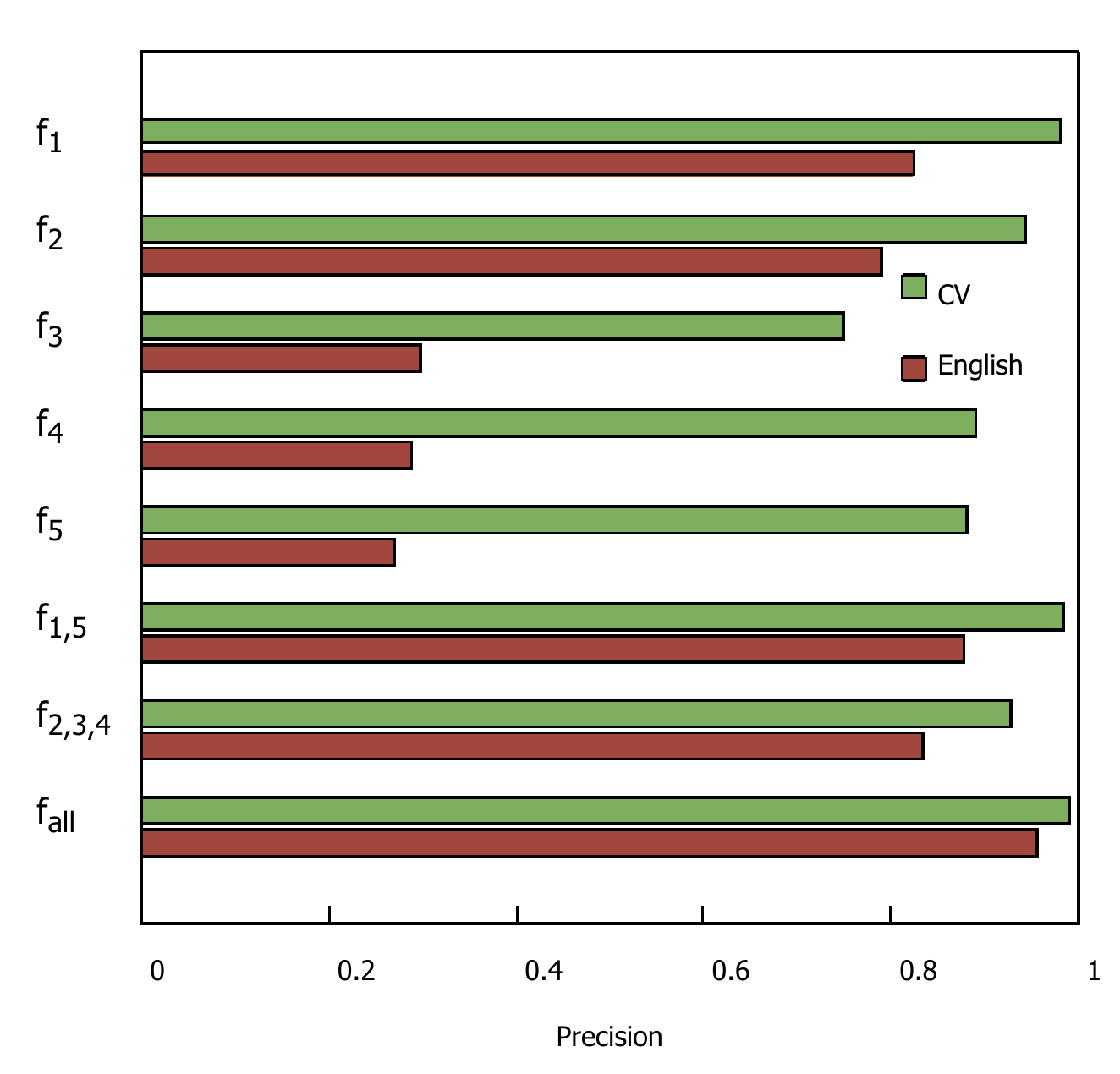}
                \caption{}
                \label{fig:precision}
        \end{subfigure}      
       ~ 
        \begin{subfigure}[b]{0.33\textwidth}
                \centering
                \includegraphics[width=\textwidth]{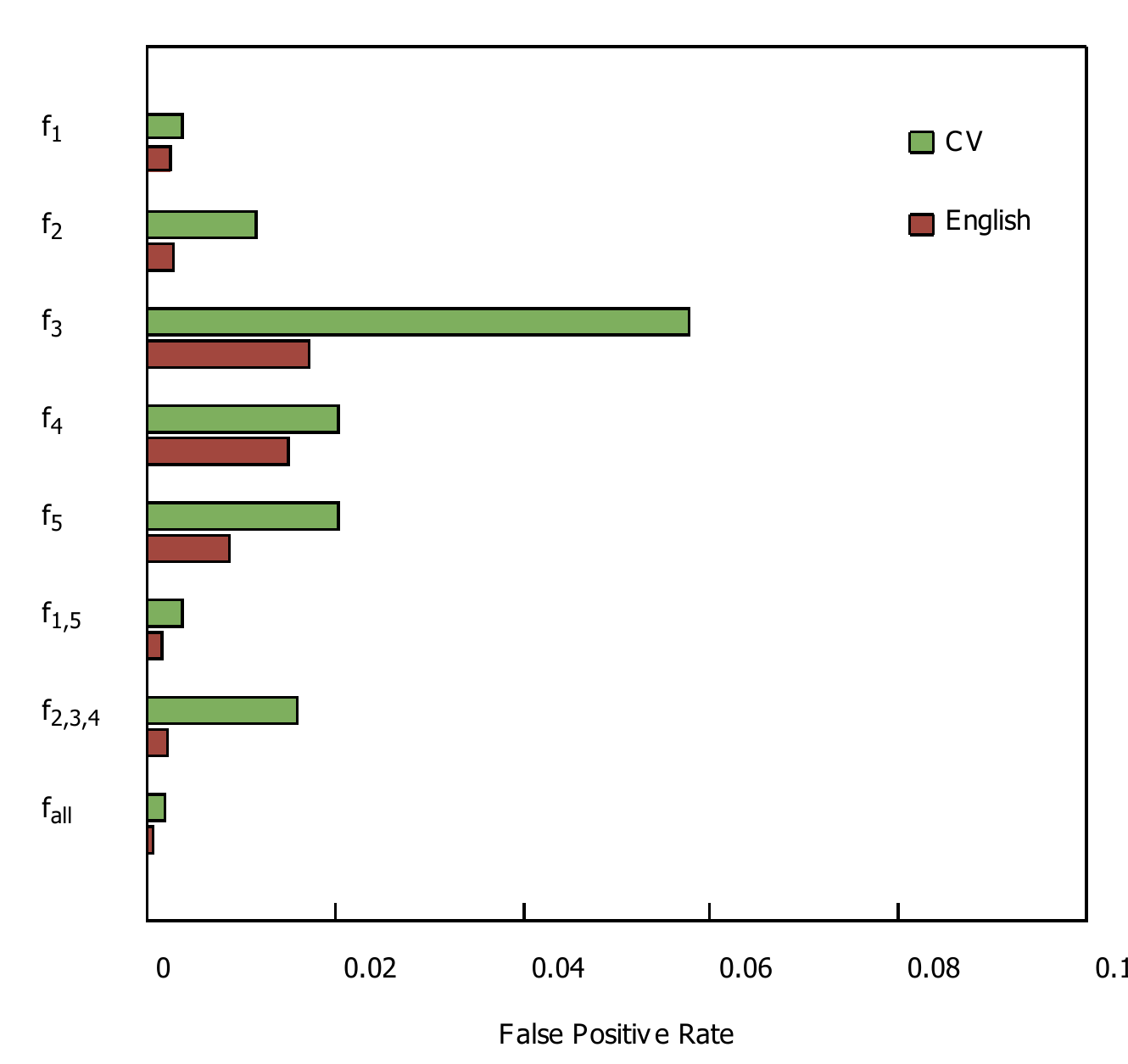}
                \caption{}
                \label{fig:fprate}
        \end{subfigure} 
         ~ 
        \caption{Accuracy evaluation results for each feature set, including (a) Recall, (b) Precision, (c) False Positive Rate}\label{fig:main_fp_eval}
\end{figure*}
\else
\fi

\subsection{Phishing Webpage Classification}
\label{subsec:exp_classif}

We now present detailed evaluation of our phishing detection method. We focus on three primary aspects of 
classification performance. First is \emph{accuracy} which entails precision, recall and false positive rate. Second is \emph{Receiver Operating Characteristic (ROC)}, which shows the change of false positive rate with respect to true positive rate. Third is \emph{scalability} where we evaluate how accuracy changes as we scale from small to large test datasets. 
We evaluate the performance of our method across six different languages so as to demonstrate its language independence. 
\iffeateval
We also evaluate each feature set in Table~\ref{tab:feature_type} independently in order to provide insights into the type of features that are most or least valuable in phishing detection.
Two scenarios are used  for evaluation. The first (\textit{scenario1}) is a 5 fold cross-validation on \textit{legTrain} and \textit{phishTrain} datasets. The second (\textit{scenario2}) 
consists of the same learning stage on \textit{legTrain} and \textit{phishTrain} (5,567 webpages), being the oldest captured datasets. Prediction is based on \textit{phishTest} and each individual language-specific test dataset of legitimate URLs. 
\else
The evaluation scenario for all languages 
consists of the same learning stage on \textit{legTrain} and \textit{phishTrain} (5,567 webpages), being the oldest captured datasets. Prediction is based on \textit{phishTest} and each individual language-specific test dataset of legitimate URLs. 
\fi

\begin{table}[th]
\caption{Detailed accuracy evaluation for six languages} \centering
	\begin{tabular}{l l l l l l }
	\textbf{Language}	& \textbf{Pre.} & \textbf{Recall}  & \textbf{$F_1$-score}  &\textbf{FP Rate}  & \textbf{AUC} \\ \hline
English & 0.956 &	0.958 &		0.957 &		0.0005	 &	0.999 \\
French & 0.970 &	0.958 &		0.964 &		0.0036	 &	0.997 \\
German &  0.981 &	0.958 &		0.970 &		0.0022 &		0.998 \\
Portuguese &  0.967	 &	0.958	& 0.962 &		0.004	 &	0.997 \\
Italian &  0.982	 &	0.958	 &	0.970	 &	0.0021 &	0.998 \\
Spanish &  0.982	 &	0.958	 &	0.970	 &	0.0021	 &	0.998 \\
		\label{tbl:main-eval-table}
 \end{tabular}
\end{table}

\iffeateval
\begin{table*}[th]
\caption{Detailed accuracy evaluation for different feature sets} \centering
	\begin{tabular}{ p{2.3cm} p{2.2cm}  p{1.1cm}  p{1.1cm}  p{1.1cm} p{1.1cm}   p{1.1cm}   p{1.1cm}   p{1.1cm}   p{1.1cm}   }
	Scenario	& Metrics & \multicolumn{8}{c}{Categories of feature sets} \\
 &  &   $f_1$ & $f_2$ & $f_3$  & $f_4$ & $f_5$ & $f_{1,5}$  & $f_{2,3,4}$ &\textbf{$f_{all}$}  \\
		\hline
 &	Precision & \textbf{\textcolor{green}{0.982}} & 0.943 &  \textbf{\textcolor{red}{0.747}} &  0.891 &  0.880 &  0.983 &  0.928 &\textbf{0.991} 	 \\
Cross-validation & Recall & \textbf{\textcolor{green}{0.932}} & 0.852 & 0.750 & 0.730 & \textbf{\textcolor{red}{0.656}} & 0.948 & 0.901 &\textbf{0.957} 	 \\
(\textit{legTrain}/& F1-score & 0.957 & 0.896 & 0.749 & 0.803 & 0.752 & 0.966 & 0.915 & \textbf{0.974}	 \\
\textit{phishTrain})& FP Rate & \textbf{\textcolor{green}{0.003}} & 0.011 & 0.057 & 0.020 & 0.020 & 0.003 & 0.015 & \textbf{0.001}	 \\ 
& AUC & 0.996 & 0.989 & 0.958 & 0.970 & 0.944 & 0.997 & 0.991 & \textbf{0.998}	 \\ \hline
	\multirow{5}{*}{English} &Precision & \textbf{\textcolor{green}{0.823}} & 0.790 & 0.296 & 0.288 & \textbf{\textcolor{red}{0.268}} & 0.879 & 0.832 &\textbf{0.956} 	 \\
& Recall &\textbf{\textcolor{green}{0.961}} & 0.858 & 0.601 & 0.502 & \textbf{\textcolor{red}{0.260}} & 0.958 & 0.847 & \textbf{0.958}	 \\
& F1-score & 0.887 & 0.823 & 0.397 & 0.367 & 0.264 & 0.917 & 0.840 & \textbf{0.957}	 \\
& FP Rate &  \textbf{\textcolor{green}{0.002}} &  \textbf{\textcolor{green}{0.002}} & 0.017 & 0.015 & \textbf{\textcolor{green}{0.008}} & 0.001 & 0.002 & \textbf{0.0005} \\ 
& AUC & 0.997 & 0.995 & 0.974 & 0.962 & 0.899 & 0.997 & 0.994 & \textbf{0.999}	 \\ 
		\label{main-eval-table}
 \end{tabular}
\end{table*}
\fi

\iffeateval
\subsubsection{Accuracy}

\textbf{Evaluation across languages:} The detailed evaluation results for precision, recall and false positive rate, using legitimate datasets of six different languages with \textit{scenario2} are shown in Table~\ref{tbl:main-eval-table}. These values were obtained by setting the discrimination threshold of Gradient Boosting to $0.7$. 
In this table, we see that our method achieves significantly high precision for all languages (0.95--0.98). This holds for recall as well (around 0.95). Hence, the $F_1$-score, which is the harmonic mean of precision and recall, is also significantly high  (0.95--0.97). The false positive rate is significantly low, i.e., in the range of 0.0005--0.004, across all languages. 
\else
\noindent\textbf{Accuracy:}
The detailed evaluation results for precision, recall and false positive rate, using legitimate datasets of six different languages are shown in Table~\ref{tbl:main-eval-table}. These values were obtained by setting the discrimination threshold of Gradient Boosting to $0.7$, which favors the prediction of legitmate webpages  ($ \left[0, 0.7 \right[ $) over phishs ($ \left[0.7, 1 \right] $). 
In this table, we see that our method achieves significantly high precision for all languages (0.95--0.98). This holds for recall as well (around 0.95). Hence, the $F_1$-score, which is the harmonic mean of precision and recall, is also significantly high (0.95--0.97). The false positive rate is significantly low, i.e., in the range of 0.0005--0.004, across all languages.
\fi

\begin{figure}[th]
                \centering
                \includegraphics[width=0.49\textwidth]{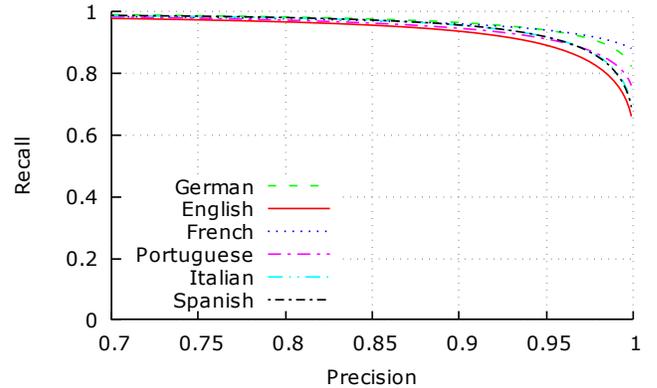}
                \caption{Precision vs recall evaluation}
                \label{fig:precision-vs-recall}
\end{figure} 

\begin{figure}[th]
                \centering
                \includegraphics[width=0.49\textwidth]{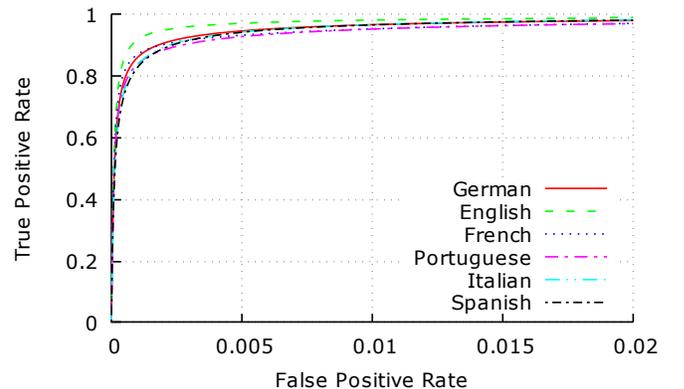}
                \caption{ROC evaluation results for six languages}
                \label{fig:roc-lang}
\end{figure} 

In many large-scale, real-world scenarios (especially in web security domain),  a machine learning model is considered usable only if it achieves high precision (e.g., 0.9 or 0.95) with significant recall (e.g., 0.5 or 0.6) \cite{singh:2012:large}. In order to test our method against this criterion, we evaluated how recall of the proposed method changes with precision by varying the discrimination threshold from 0 to 1. The result is shown in Fig.~\ref{fig:precision-vs-recall} where we see that when the precision is higher than 0.9, the recall for all languages is significantly high and is always in the range of 0.64--0.98. This shows that from the accuracy perspective, our method is readily applicable  in large-scale, multi-lingual business scenarios.

\iffullversion
\textbf{Evaluation across feature sets:} For evaluating each feature set individually, we experimented \textit{scenario1} (cross-validation) and \textit{scenario2} with the English dataset (English). The latter scenario corresponds to a real world scenario where the ratio of legitimate webpages to phishs is 85/1 as it has been observed while analyzing real world traffic (90/1) in previous work \cite{whittaker:2010:large}.
The  evaluation results for both these experiments using a discrimination threshold set to $0.7$, with respect to precision, recall and false positive rate of each feature set are shown in Fig.~\ref{fig:main_fp_eval}.  

In Fig.~\ref{fig:precision}, we see that, out of all feature sets (i.e., $f_1$,$f_2$,$f_3$,$f_4$,$f_5$), $f_1$ yields the highest precision whereas  $f_3$ or  $f_5$ yield the lowest precision. But, note that for English dataset, the precision of $f_1$ is 0.82 which is not as high as it is for cross-validation. In this case, the rest of the feature sets (i.e., $f_2$, $f_3$, $f_4$,  and $f_5$) become significantly important, as combining them together in $f_{all}$ increases the precision from 0.82 (achieved by $f_1$) to 0.95 (achieved by $f_{all}$).   
On similar lines, we see in Fig.~\ref{fig:recall} that out of  all individual feature sets, $f_1$ performs the best in terms of recall as well whereas $f_5$ performs the worst. 
The detailed results can also be seen in highlighted text in Table~\ref{main-eval-table}. We can see that the overall recall of  the proposed method, achieved by using $f_{all}$, is significantly high and higher than 0.95.

Fig.~\ref{fig:fprate} shows the false positive rate of each feature set wherein we see that the overall false positive rate of the proposed method (obtained by using feature set $f_{all}$) is very low, i.e., in the range of 0.0005--0.001, which illustrates the practical applicability of the method in real-world scenarios. It is of note that for a large dataset like English, the false positive rate of the method is as low as 0.0005. This can be mainly attributed to feature set $f_1$ and $f_2$, both of which yield very low false positive rate. 
This is also shown in Table~\ref{main-eval-table} where we see that, just like for precision and recall, feature set $f_3$ individually yields the least performance which indicates that features based on usage of starting and landing mld, if used by themselves, are not good indicators of a phishing website, and must be used in conjunction with other features.
\fi

\iffeateval
\begin{figure*}[th]
        \begin{subfigure}[b]{0.24\textwidth}
                \centering
                \includegraphics[width=\textwidth]{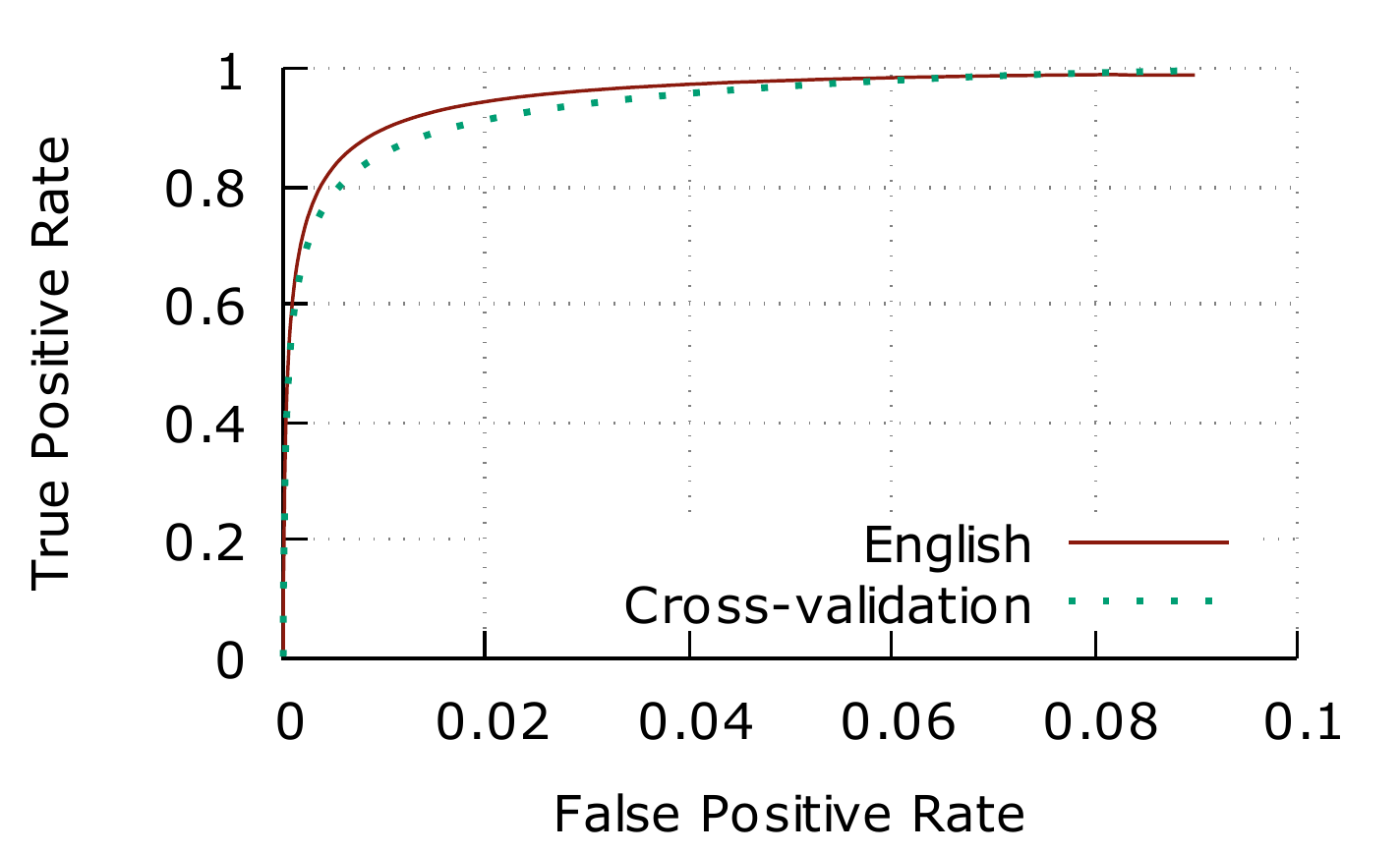}
                \caption{}
                \label{fig:roc_f1}
        \end{subfigure}%
        ~ 
        \begin{subfigure}[b]{0.24\textwidth}
                \centering
                \includegraphics[width=\textwidth]{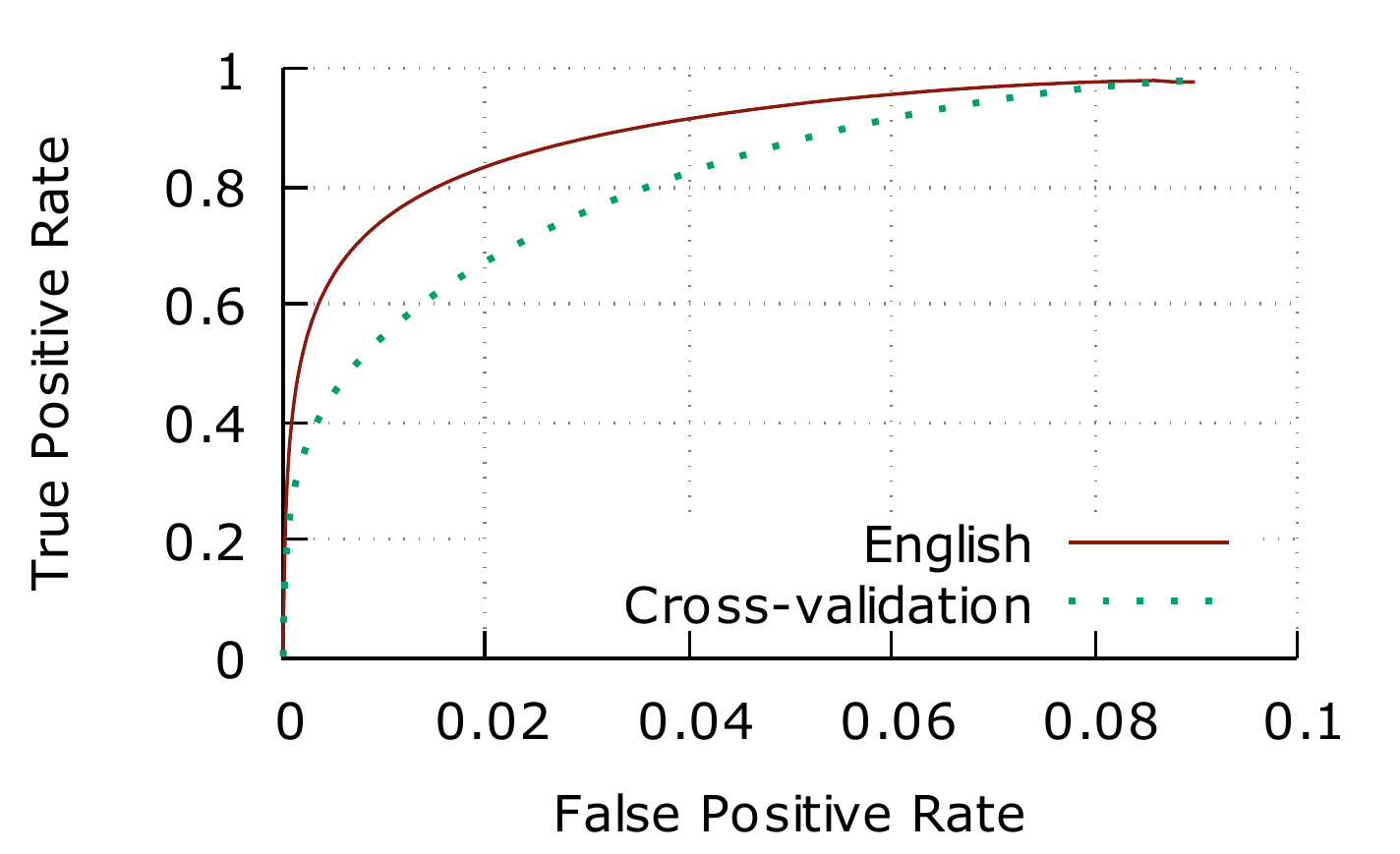}
                \caption{}
                \label{fig:roc_f2}
        \end{subfigure}      
       ~ 
        \begin{subfigure}[b]{0.24\textwidth}
                \centering
                \includegraphics[width=\textwidth]{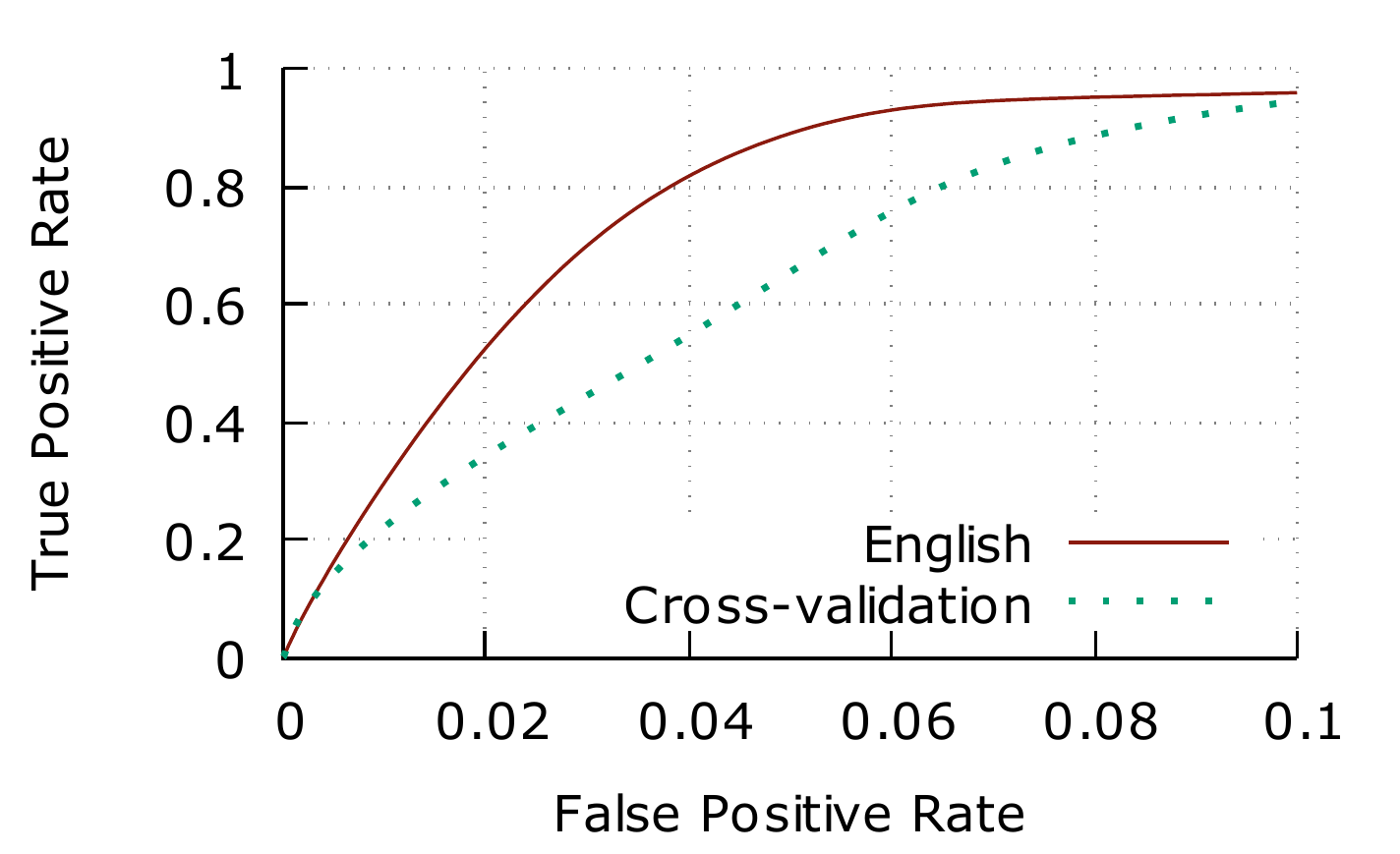}
                \caption{}
                \label{fig:roc_f3}
        \end{subfigure} 
         ~ 
		 \begin{subfigure}[b]{0.24\textwidth}
                \centering
                \includegraphics[width=\textwidth]{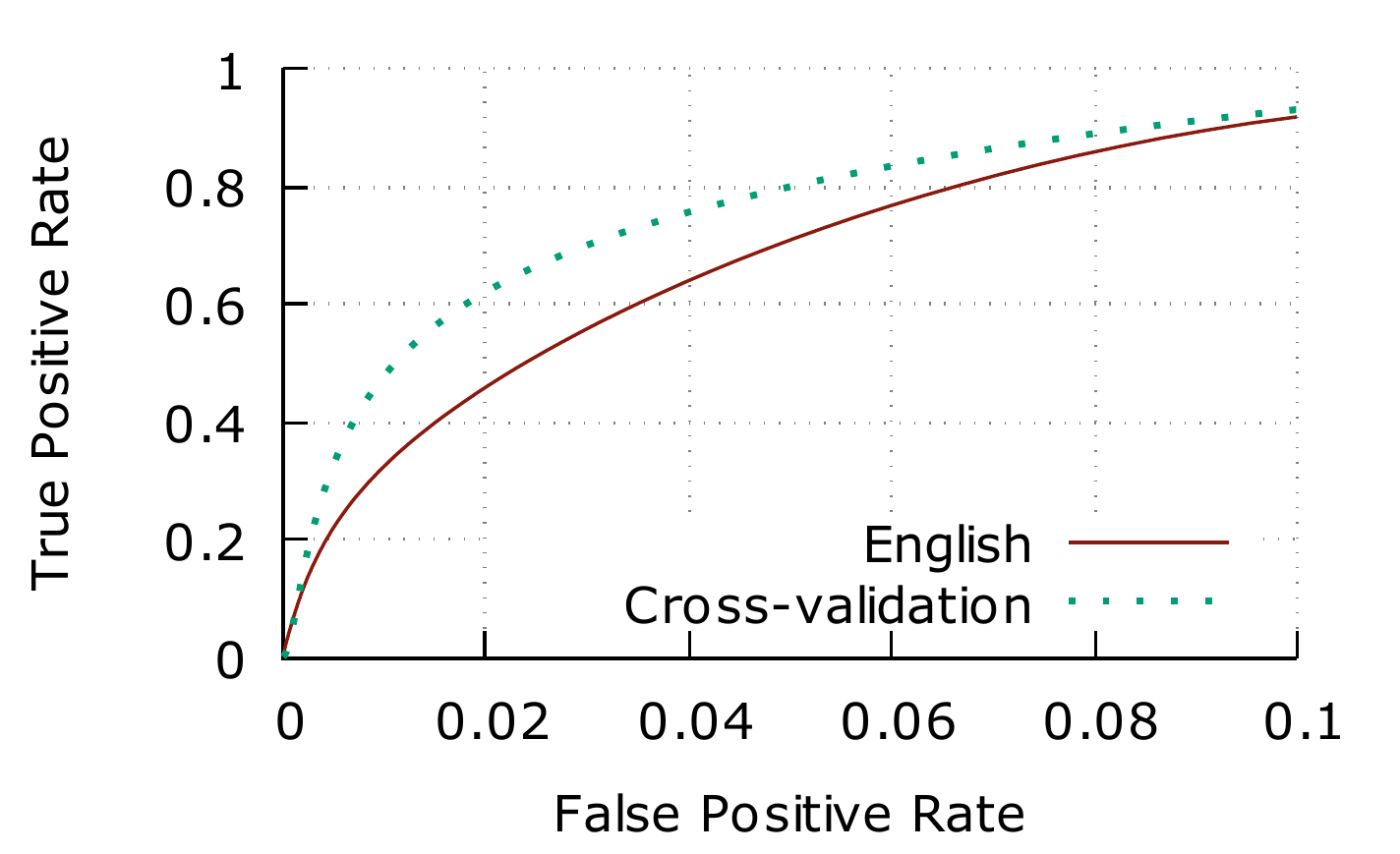}
                \caption{}
                \label{fig:roc_f4}
        \end{subfigure}
        
        \begin{subfigure}[b]{0.24\textwidth}
                \centering
                \includegraphics[width=\textwidth]{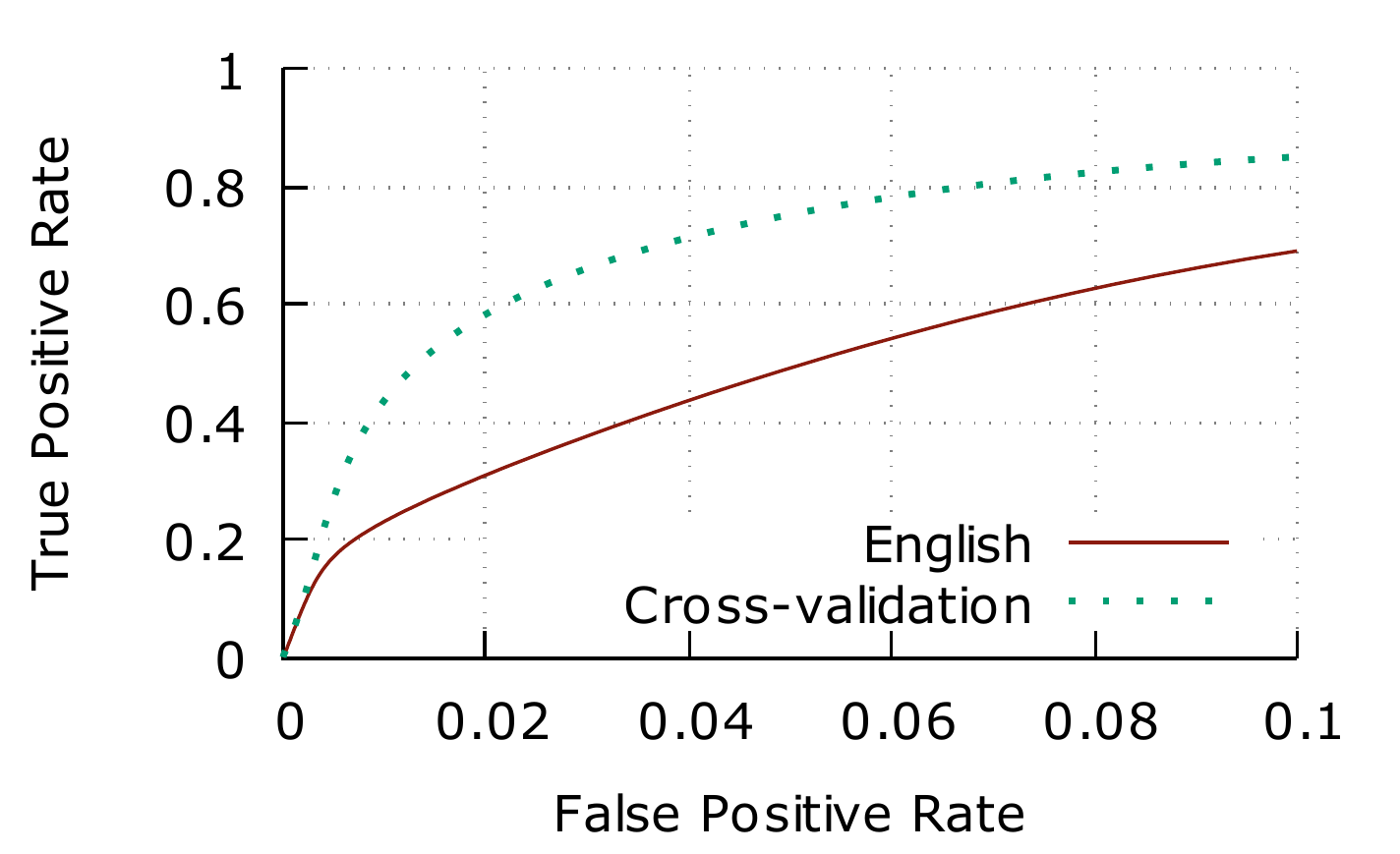}
                \caption{}
                \label{fig:roc_f5}
        \end{subfigure}      
      		~  \begin{subfigure}[b]{0.24\textwidth}
                \centering
                \includegraphics[width=\textwidth]{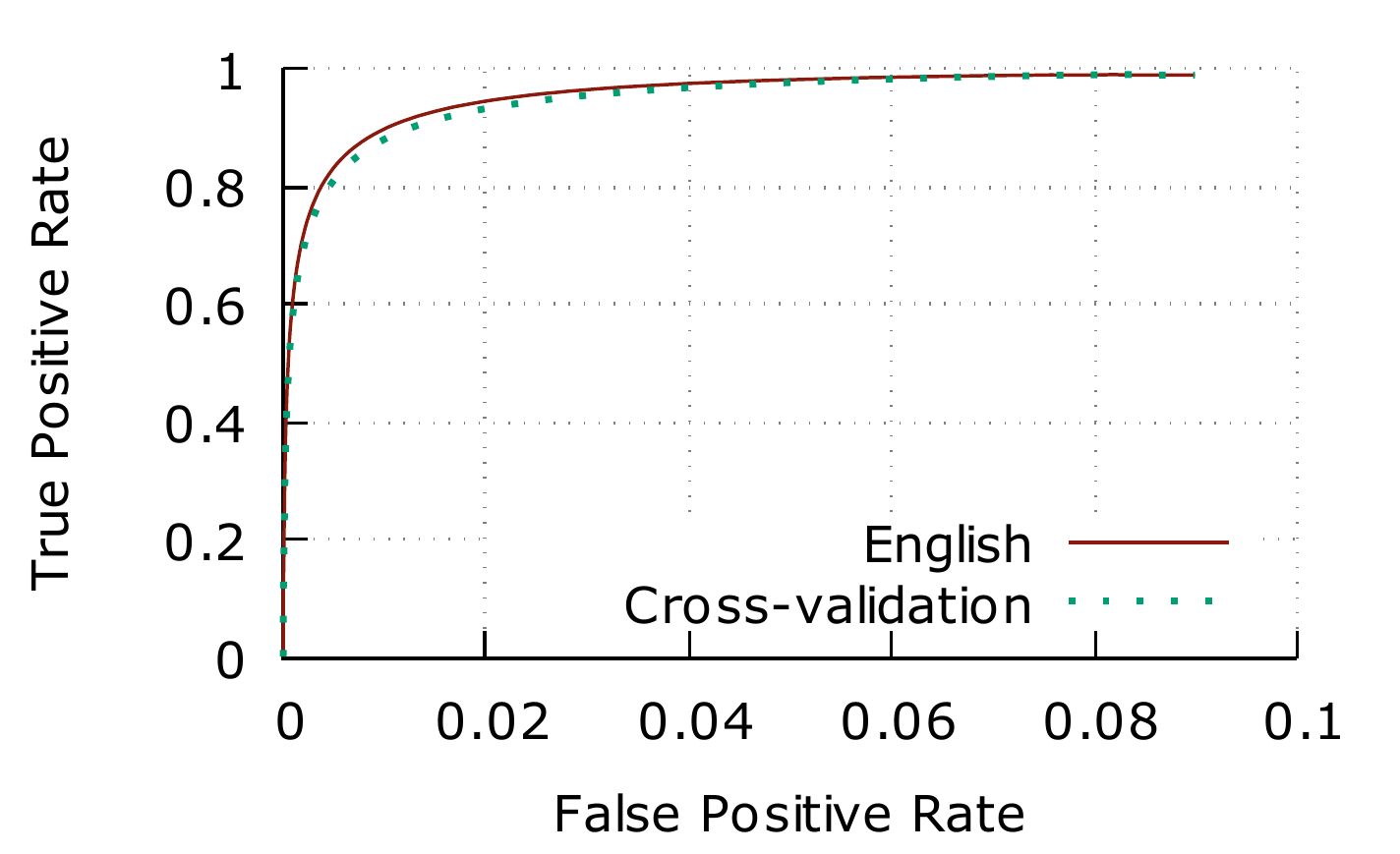}
                \caption{}
                \label{fig:roc_f1to5}
        \end{subfigure} 	
        		~\begin{subfigure}[b]{0.24\textwidth}
                \centering
                \includegraphics[width=\textwidth]{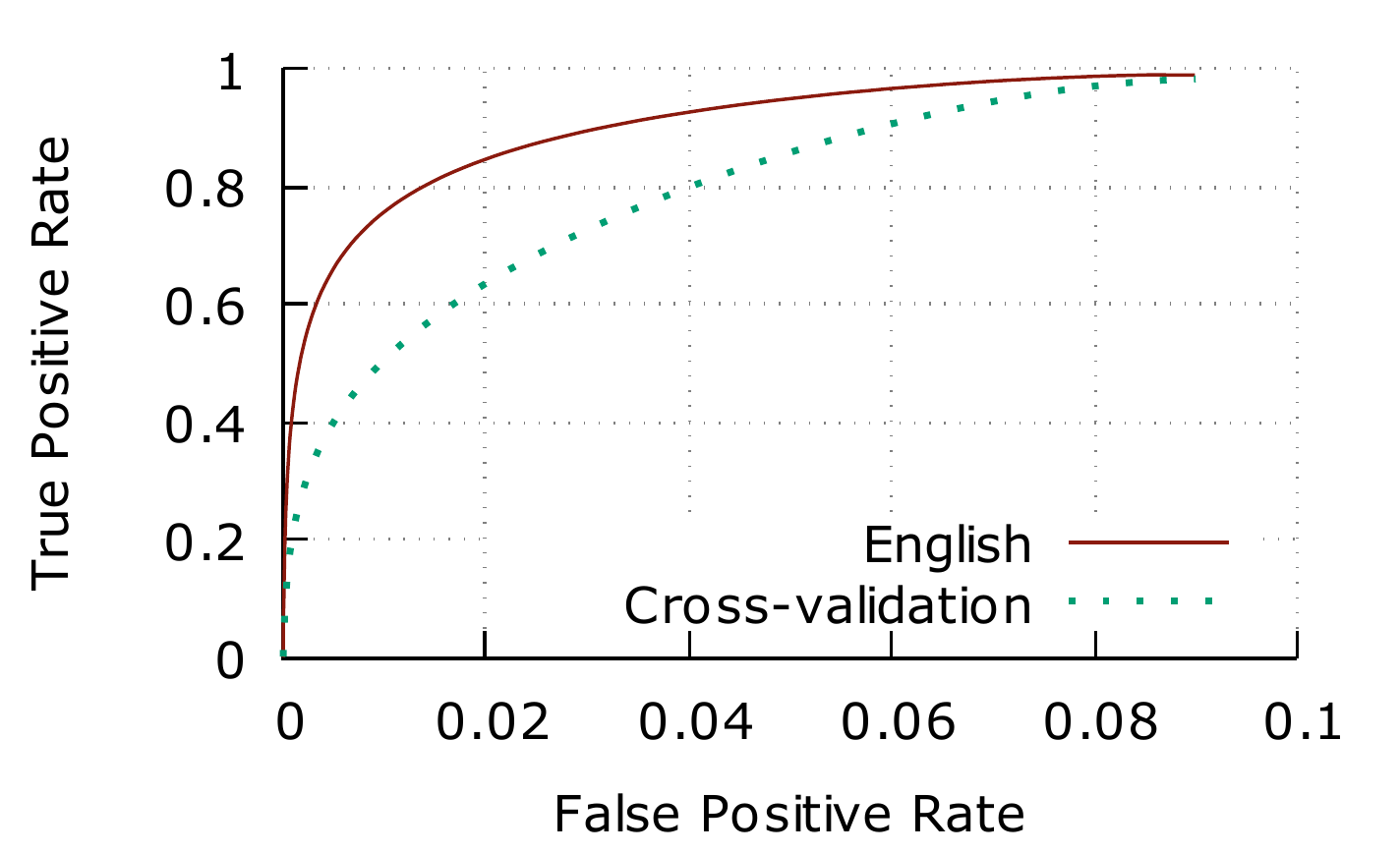}
                \caption{}
                \label{fig:rocf234}
        \end{subfigure} 	
  		~      \begin{subfigure}[b]{0.24\textwidth}
                \centering
                \includegraphics[width=\textwidth]{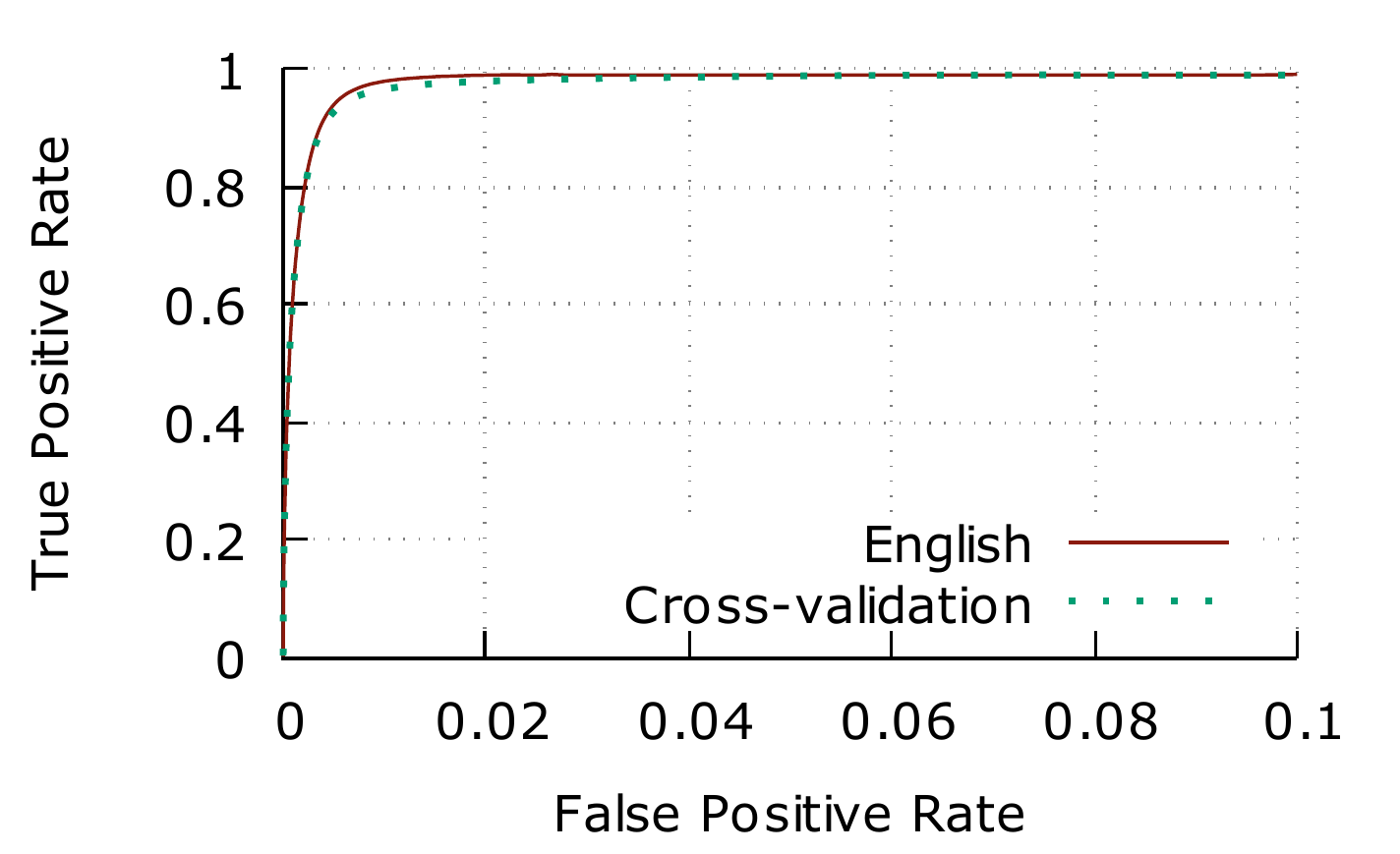}
                \caption{}
                \label{fig:roc_all}
        \end{subfigure} 						 

        \caption{ROC evaluation for different feature sets: (a) $f_1$, (b) $f_2$, (c) $f_3$, (d) $f_4$, (e) $f_5$, (f) $f_{1,5}$, (g) $f_{2,3,4}$, (h) $f_{all}$}\label{fig:roc}
\end{figure*}
\fi

\iffeateval

\textbf{Evaluation across feature sets:} For evaluating each feature set individually, we experimented \textit{scenario1} (cross-validation) and \textit{scenario2} with the English dataset (English). The latter scenario corresponds to a real world scenario where the ratio of legitimate webpages to phishs is 85/1 as it has been observed while analyzing real world traffic (90/1) in previous work \cite{whittaker:2010:large}.

Detailed performance results, for a discrimination threshold of $0.7$, are provided in Table~\ref{main-eval-table} where we highlight the maximum and minimum precision, recall and false positive rate for each scenario. We see that out of all individual feature sets (i.e., $f_1$,$f_2$,$f_3$,$f_4$,$f_5$), $f_1$ yields the maximum precision, whereas  $f_3$ or  $f_5$ yield the minimum precision. 
But, note that for English dataset, the precision of $f_1$ is 0.82 which is not as high as it is for cross-validation. In this case, the rest of the feature sets (i.e., $f_2$, $f_3$, $f_4$,  and $f_5$) become significantly important, as combining them together in $f_{all}$ increases the precision from 0.82 (achieved by $f_1$) to 0.95 (achieved by $f_{all}$).  
On similar lines, we see that out of all individual feature sets, $f_1$ performs the best in terms of recall as well whereas $f_5$ performs the worst.
The overall false positive rate of the proposed method (obtained by using feature set $f_{all}$) is very low (0.0005--0.001), which illustrates the practical applicability of the method in real-world scenarios. Note that for a large dataset like English, the false positive rate of the method is as low as 0.0005. This can be mainly attributed to feature set $f_1$ and $f_2$, both of which yield very low false positive rate.
\fi

\iffeateval
\subsubsection{ROC}
\else
\noindent\textbf{ROC:}
\fi
Another  metric for predictive performance of the proposed method is ROC and corresponding AUC (Area Under the Curve).
\iffeateval
Along the lines of accuracy evaluation, we examine the ROC and AUC metrics across all languages, and across individual feature sets. 
\else
Along the lines of accuracy evaluation, we examine the ROC and AUC metrics across all languages.
\fi
\iffeateval
\textbf{Evaluation across languages:} 
The objective of ROC evaluation is to examine the increase in false positive rate with the increase in true positive rate while varying the discrimination threshold of the classifier. The evaluation results using \textit{scenario2} for all languages are shown in Fig.~\ref{fig:roc-lang}. We see that, at the significantly high true positive rate of 0.9, the false positive rate for all languages is less than 0.008 which is considered quite low. As the true positive rate increases to around 0.95, the false positive rate does not increase much.  Even at true positive rate of 0.98, the false positive rate stays substantially low at 0.02. In line with these results, the AUC is around 0.999 for all languages, as shown in Table~\ref{tbl:main-eval-table}. It may be noted that these results are consistent across all languages, which is very desirable in a multi-lingual phishing detection scenario. \eat{It is only when the true positive rate increases to 0.9999 that the false positive rate goes up to 0.1. But, in real-world phishing detection scenarios, generally}
\else
The objective of ROC evaluation is to examine the increase in false positive rate with the increase in true positive rate while varying the discrimination threshold of the classifier. The evaluation results for all languages are shown in Fig.~\ref{fig:roc-lang}. We see that, at the significantly high true positive rate of 0.9, the false positive rate for all languages is less than 0.008 which is considered quite low. As the true positive rate increases to around 0.95, the false positive rate does not increase much.  Even at true positive rate of 0.98, the false positive rate stays substantially low at 0.02. In line with these results, the AUC is around 0.999 for all languages, as shown in Table~\ref{tbl:main-eval-table}. Note that these results are consistent across all languages, which is very desirable in a multi-lingual phishing detection scenario. \eat{It is only when the true positive rate increases to 0.9999 that the false positive rate goes up to 0.1. But, in real-world phishing detection scenarios, generally}
\fi

\iffeateval
\textbf{Evaluation across feature sets:} For ROC evaluation of individual feature sets, we used the same experimental setup as we did for accuracy evaluation across feature sets, i.e., \textit{scenario1} (cross-validation) and \textit{scenario2} with English dataset (English). 
The results are shown in Fig.~\ref{fig:roc}. We see that, similar to accuracy evaluation results, feature $f_1$ performs the best as it has maximum area under ROC curve in both the scenarios whereas features $f_3$ and $f_5$ perform the least in both the scenarios. This indicates that in this type of scenario, features based on usage of starting and landing \textit{mld} (i.e., $f_3$) and features based on webpage content (i.e., $f_5$), if used by themselves, are not good indicators of a phishing website, and must be used in conjunction with other features.
\fi

\begin{figure}[th]
                \centering
                \includegraphics[width=0.49\textwidth]{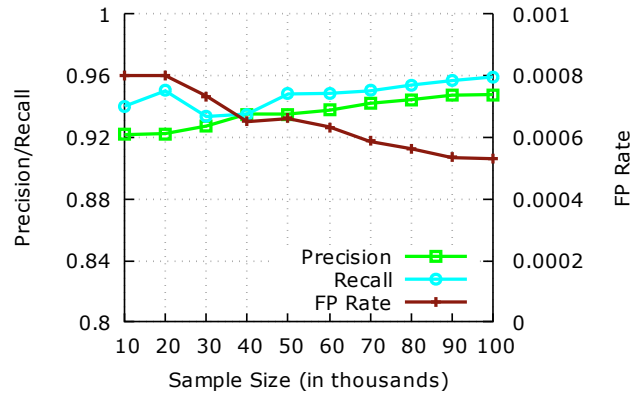}
                \caption{Performance vs the scale of data}
                \label{fig:scalability-eval}
\end{figure}

\iffeateval
\subsubsection{Scalability}
\else
\noindent\textbf{Scalability:}
\fi
We now examine the effect of scale on the predictive performance of our method, i.e., if the size of the test dataset increases considerably with time, then what effect does it have on the precision, recall and false positive rate?
\iffeateval
This evaluation is especially important for verifying the practical usability of the method because in real-world, business scenarios,  machine learning models are trained once and then tested or applied on large, and often streaming, sets of data which rapidly increase with time. In such problem scenarios, the learnt model must be able to scale well with the increase in the size of datasets.

In order to perform this scalability evaluation, we use \textit{scenario2} (5,567 training instances) and initialize our test set with 10,000 legitimate and 100 phishing examples extracted randomly from the English dataset and \textit{phishTest} respectively. Thereafter, we iteratively increment the size of the test set by 10,000 legitimate webpages and 100 phishs randomly picked from the remaining instances of the English dataset and \textit{phishTest}.
\else
We initialize our test set with 10,000 legitimate and 100 phishing examples extracted randomly from the English dataset and \textit{phishTest} respectively. Thereafter, we iteratively increment the size of the test set by 10,000 legitimate webpages and 100 phishs randomly picked from the remaining instances of the English dataset and \textit{phishTest}.
\fi

\iffullversion
This helps us in examining how the predictive performance of our phishing detection model changes as we increase the size of test set from 10,100 to 101,000. 
\fi
The results are shown in Fig.~\ref{fig:scalability-eval}, where we see that precision as well as recall increase with scale, whereas the false positive rate decreases. This indicates that the increase in the number of errors (i.e., false positives and false negatives) is significantly less than the increase in the size of test set, which causes the increase in overall precision and recall and decrease in the false positive rate as we scale up the size of test set. This kind of predictive performance on large test set, while learning a model from a small training set,  is exactly what is required in a desirable machine learning model that is deemed fit for usage in large-scale practical scenarios. 

\begin{table}[tbh]
\centering
\begin{tabular}{l r r r}

 & \textbf{Median} & \textbf{Average} & \textbf{StDev} \\ \hline
Webpage scraping & 12787 & 12798 & 4898 \\
Loading data & 1 & 2 & 2 \\ 
Features extraction & 890 & 1282 & 1586 \\ 
Classification & $<1$ & $<1$ & $<1$ \\ \hline
Total (no scraping) & 891 & 1283 & 1588

\end{tabular}
\caption{Processing time (milliseconds)}
\label{tab:time}
\end{table}

Table~\ref{tab:time} depicts the median, average and standard deviation of the time taken by each operation involved in phishing webpage classification on a laptop with 2.7GHz Intel Core i5 processor and 16GB memory. We can see that most of the time is dedicated to webpage scraping, which is not part of the classification process in the case of client-side implementation. Except that, the total median classification time is 891 milliseconds, showing that the system is able to render a decision in less than 1 second. These performance figures are based on a standalone Python prototype. Subsequently, we also implemented an optimized JavaScript version as a browser add-on which exhibits better performance characteristics~\cite{armano:2016:real}.


\subsection{Target Identification}
\label{subsec:target_eval}

To assess the performance of target identification, we used the 600 phishing webpages of \textit{phishBrand}.
Since target identification can provide up to three candidate targets for an analyzed webpage, we evaluate the likelihood of the correct target being part of top-1, top-2 and top-3 results provided by our scheme.
Table \ref{tab:target_results} presents the count of correctly identified targets, unknown targets and missed targets considering these three sets. The last column gives the success rate of each method. 
The 17 pages with unknown target corresponds to webpages including only some input fields and no hint about the target. We were not able to infer the target with manual analysis. 
\iffeateval
We assume that links to these webpages were provided in phishing emails that contained the information about the target.
\fi
These webpages with unknown targets are thus included in the computing of the success rate.
The accuracy of identifying the correct target (top-1) is 90.5\%. If the criteria is identifying the correct target among a possible set of 3 (top-3) then the accuracy increases to 97.3\%.
\iffeateval
311 phishing webpages had only one identified potential target (top-1) and no alternative targets.
\fi
These results are comparable to the best state of the art method for target identification \cite{liu:2012:anti} that reaches a 92.1\% success rate.
\iffeateval
However, our method is significantly faster to compute, since it it does not require crawling additional websites to infer the target of a given phish.
\fi

\begin{table}[tbh]
\caption{Target identification results} \centering
\begin{tabular}{l c c c c}

\textbf{Targets} & \textbf{Identified} & \textbf{Unknown} & \textbf{Missed} & \textbf{Success rate} \\ \hline
top-1 & 526 & 17 & 57  & 90.5\%\\
top-2 & 558 & 17  & 25 & 95.8\%\\
top-3 & 567 & 17 & 16 & 97.3\%\\
\end{tabular}
\label{tab:target_results}
\end{table}

\iffeateval
To see how the target identification system can complement our phishing detection system we fed the former with misclassified legitimate webpages identified in Section \ref{subsec:exp_classif} when assessing phishing detection in \textit{scenario2} with the English dataset. 53 out of 100,000 legitimate webpages were misclassified. The target identification system identified four of these as phish with an identified target. 10 were considered as suspicious (no target identified and no legitimate confirmation) and 39 were confirmed as legitimate.
\else
To see how the target identification system can complement our phishing detection system we fed the former with misclassified legitimate webpages identified in Section \ref{subsec:exp_classif} when assessing phishing detection with the English dataset. 53 out of 100,000 legitimate webpages were misclassified. The target identification system identified four of these as phish with an identified target. 10 were considered as suspicious (no target identified and no legitimate confirmation) and 39 were confirmed as legitimate.
\fi

Considering these results, using the target identification in a second step for instances identified as phishs by the phishing detection system can be beneficial. On the \textit{English} dataset, it would reduce the false positive rate to 0.0001, which is equal to the best state-of-the-art phishing detection system \cite{whittaker:2010:large}. However, according to accuracy in target identification (97.3\%) it would as well reduce the number of identified phishs while keeping precision and recall over 0.90.
\section{Discussion}

\iffullversion
Analyzing the design and the evaluation results of phishing classification and target identification, we draw some conclusion and discuss the limitations of this technique as well as potential adversarial attacks.
\fi

\subsection{Relevance of Feature Sets}

We have seen in Section~\ref{subsec:exp_classif} that our features set yielded results that outperform previous work. The main reason for this improvement is the new separation scheme applied to data sources related to their level of control and constraints (Section~\ref{subsec:phisher_limitation}). 
\iffeateval
This is evident from the performances of the feature set $f_1$ that comprises URL related features separated accordingly to constraint and control considerations.
The consistency checking in term usage represented by $f_2$ showed very good performances as well and the combination of all feature sets ($f_{all}$) yielded comparable results to the best existing techniques \cite{whittaker:2010:large}, while relying on less features and training data.
\else
Analysing the weight of each features in the learnt classification model we observed that individually, features belonging to $f_2$, were the most relevant. We omit details about feature analysis for lack of space, see our research report \cite{arXiv} for details. This explains the reason why we obtained comparable results to the best existing techniques \cite{whittaker:2010:large}, while relying on less features and training data.
\fi

In addition, we assessed that our feature set meets the requirements introduced in Section~\ref{subsec:requirements}. It has good generalizability being able to learn a classification model from few thousand instances and accurately predicting the class of 100,000+ unknown webpages. It is adaptable and language/brand independent achieving comparable performances across different languages. It is usable as it does not rely on online access to centralized information and is fast to render a decision with a median processing time lower than 1 second.
 
\subsection{Limitations}

The main strength of our technique, its language independence, is though its main weakness. 
\iffeateval
We did not want to rely on any dictionary to extract terms. 
\fi
We chose to split strings according to any characters that are not part of the English dictionary and to only consider terms composed of at least three characters to discard stop words and recurrent short terms having no meaning. This raised some issues in term distribution comparisons. Long subdomains such as \textit{theinstantexchange} or \textit{insuranceservicenow} were considered as single term. 
In contrast, short domain name string corresponding to brand and composed of separating characters (digit, hyphen, etc.) such as \textit{dl4a}, \textit{s2mr} or \textit{e-go} were split and the resulting terms were discarded as too short.
The inconsistent usage of abbreviations or acronyms like \textit{intl} for \textit{international} or \textit{pfa} for \textit{premier financial online} also had a negative impact. Similarity of synonyms cannot be inferred. Most misclassified legitimate webpages (\textgreater 50\%) had one of these characteristics. Despite these misclassifications we achieve a low false positive rate (0.0005). Many of these misclassified instances can be identified as legitimate by the target identification system.	

A second limitation relates to the identification of some empty/unavailable webpages and parked domain names as phishs. The former is explained by the lack of information contained in empty/unavailable webpages. 
\iffullversion
The text and title contents are almost empty and few outgoing links or external resources (logged links) are loaded in these pages.
\fi 
Several parked domain names are domains that have been used for malicious purposes like phishing \cite{li:2013:finding} and are thus obfuscated \textit{FQDN}s registered to trap users. Moreover, parked domain names use similar composition schemes and obfuscation techniques as phishing domains 
\cite{vissers:2015:parking} such as
\iffeateval
typosquatting \cite{agten:2015:seven,szurdi:2014:long}.
\else
typosquatting \cite{szurdi:2014:long}.
\fi 
\iffullversion
In addition to this similarity in domain names and URLs composition, parked domain names and phishing domain names have other common characteristics. 
\else
Parked domain names and phishing domain names have other common characteristics.
\fi
\iffeateval
Parked domains are involved in advertisement networks \cite{alrwais:2014:understanding} and the delivered ad content tends to be correlated with the domain name parked, for instance ads for Amazon Inc. are delivered for the \textit{RDN} \textit{amaaon.com}. 
From the point of view of our classification system, these parked pages have the same characteristics as phishing pages. 
This misclassification of unavailable and parked domain names is not of major concern since, for the former no content access is prohibited by the system since the link point empty resources. For the latter, domain parking is considered as an activity that provide very \textit{little unique content} and is considered as spam by Google \cite{google:spam}. Nevertheless, some efficient state-of-the-art techniques \cite{vissers:2015:parking} or the target identification system can be applied to discard these webpages from phishing identification.
\else
Parked domains are involved in advertisement networks \cite{alrwais:2014:understanding} and the delivered ad content tends to be correlated with the domain name parked, for instance ads for Amazon Inc. are delivered for the \textit{RDN} \textit{amaaon.com}.
From the point of view of our classification system, these parked pages have the same characteristics as phishing pages. 
This misclassification of unavailable and parked domain names is not of major concern since, for the former no content access is prohibited by the system since the link point empty resources. For the latter, domain parking is considered as spam by major Internet actors (\textit{e.g.} Google) and some efficient state-of-the-art techniques \cite{vissers:2015:parking} or the target identification system can be applied to discard these webpages from phishing identification.
\fi

A last limitation was the low accuracy observed in classification of IP-based phishing URLs. Out of 25 such URLs in \textit{phishTest}, only 19 were correctly classified rendering a lower recall (0.76) than the global recall presented by the system (\textgreater 0.95). The reason is that \textit{FQDN}s based term distributions for such URLs are empty leading to several null features. However, such URLs represent less than 2\% (41) of the URLs present in all phishing datasets and is thus not a major limitation.

Although we did not observe this in our datasets, webpages whose content is in one alphabet and URL in another may be misclassified. So far we have only tested webpages in European languages. Classifier performance on pages in other languages may be lower.


\subsection{Evasion Techniques}

As we saw, one way to evade detection is to use IP-based URLs. These are less likely to be detected by our system. However, relying on IP address rather than domain names deprives phishers from the flexibility brought by the DNS to change the hosting location of their phishing content while keeping the same link. Moreover, IP blacklisting is widely used to prevent access to malicious hosting infrastructure, so phishers would have to face other issues.

Another evasion technique is to limit the text content available in a webpage: use few external links, do not load external content and build short 
URLs \cite{maggi:2013:two}.
We observed some of these techniques actually being used individually in webpages of both phishing datasets used for evaluation. They did not impact classifier performance because even though they prevent some features from being computed, others, such as those based on title, starting/landing URL and logged links could still lead to effective detection of phishs. Simultaneous use of multiple evasion techniques may impact classifier performance.
However, using such subterfuges would impact the quality of the phishing webpage and reduce the number of victims. 

\iffullversion
Image based webpages is another technique to limit text content. While such webpages can be identified by some features of our model, it would make the identification process harder. A solution to cope with this is to use OCR to extract the text content from the webpage screenshot.
\fi

A final probable evasion technique is to use typosquatting domains and misspelled terms in the different data sources we analyze. When different but similar terms like \textit{paypal}, \textit{paypaI} or \textit{paipal} are used in different sources, our distributions comparison metric would not infer any similarity.
The classifier would thus probably conclude that the webpage is legitimate. However, the presence of references to the target would disclose the real target. In addition, misspellings may tip-off potential victims.

For target identification, the best evasion technique is not to provide any indication about the target in the webpage and rather focus on using lures in the message containing the link to the fake website. But this has two negative effects, first, the phishing webpage seems less legitimate and second, the phisher exposes himself to alternative target identification techniques applied to other content than webpages \cite{ramanathan:2013:phishing}.
\section{Related Work}
\label{sec:soa}

\iffeateval
\begin{table*}[tbh]
\centering
\begin{tabular}{l r r c c c c l l l l }
& \multicolumn{2}{c}{\textbf{Testing set}} & \textbf{Legitimate} & \textbf{Train} & \textbf{Leg} & & & & & \\
\textbf{Technique} &  \textbf{Legitimate} & \textbf{Phish} & \textbf{set} & \textbf{/Test} & \textbf{/Phish}  & \textbf{Evaluation}  & \textbf{FPR}  & \textbf{Pre.} & \textbf{Recall} & \textbf{Acc.}\\ \hline
Cantina \cite{zhang:2007:cantina} & 2,100 & 19 & English & - & \textbf{\textcolor{green}{110/1}} & no learning & \textbf{\textcolor{red}{0.03}} & \textbf{\textcolor{red}{0.212}} & 0.89 & 0.969 \\
Cantina+ \cite{xiang:2011:cantina} & 1,868 & 940 & several & 1/4 & 2/1 & old/new & \textbf{\textcolor{red}{0.013}} & 0.964 & 0.955 & 0.97 \\
Xiang \textit{et al.} \cite{Xiang:2009:hybrid} & 7,906 & 3,543 & several & - & 2/1 & no learning & \textbf{\textcolor{red}{0.019}} & 0.957 & 0.9 & 0.955 \\
Ma \textit{et al.} \cite{Ma:2009:beyond} &  15,000 & 20,500 &  DMOZ & 1/1 & \textbf{\textcolor{red}{3/4}} & cross-valid& 0.001 & \textbf{\textcolor{green}{0.998}} & 0.924 & 0.955 \\
Whittaker \textit{et al.} \cite{whittaker:2010:large} &  1,499,109 & 16,967 & several & \textbf{\textcolor{red}{6/1}} & \textbf{\textcolor{green}{90/1}} & old/new & \textbf{\textcolor{green}{0.0001}} & 0.989 & 0.915 & \textbf{\textcolor{green}{0.999}} \\
Thomas \textit{et al.} \cite{thomas:2011:design} & 500,000 & 500,000 & several & 4/1 & 1/1 & cross-valid & 0.003 & 0.961 & \textbf{\textcolor{red}{0.734}} & \textbf{\textcolor{red}{0.866}} \\
Ramesh \textit{et al.} \cite{ramesh:2014:efficious} & 1,200 & 3,374 & \textbf{\textcolor{red}{top Alexa}} & - & \textbf{\textcolor{red}{1/3}} & no learning & \textbf{\textcolor{red}{0.005}} & \textbf{\textcolor{green}{0.998}}  & \textbf{\textcolor{green}{0.996}} & 0.996 \\
Chen \textit{et al.} \cite{Chen:2014:anti} & 404 & 1,945 & \textbf{\textcolor{red}{top Alexa}} & \textbf{\textcolor{red}{9/1}} & \textbf{\textcolor{red}{1/5}} & cross-valid & \textbf{\textcolor{red}{0.007}} & 0.992& \textbf{\textcolor{green}{1}}  & 0.994 \\ \hline
Our method & 100,000 & 1,216 & English & \textbf{\textcolor{green}{1/18}} & \textbf{\textcolor{green}{85/1}} & old/new & \textbf{\textcolor{green}{0.0005}} & 0.956  & 0.958 & \textbf{\textcolor{green}{0.999}} \\
Our method & 150,000 & 1,216 & several & \textbf{\textcolor{green}{1/27}} & \textbf{\textcolor{green}{125/1}} & old/new & 0.001 & 0.857  & 0.958 & \textbf{\textcolor{green}{0.998}} \\
Our method & 4,531 & 1,036 & English & 4/1 & 4/1 & cross-valid& 0.001 & \textbf{\textcolor{green}{0.991}} & 0.957  & 0.99 \\

\end{tabular}
\caption{Phishing detection system performances comparison}
\label{tab:compare}
\end{table*}

\eat{
\begin{table*}[tbh]
\centering
\begin{tabular}{l r r r r r r l l l l }

\textbf{Technique} &  \textbf{Legit} & \textbf{Phish} & \textbf{Legit} & \textbf{Phish} & \textbf{Leg set} & \textbf{evaluation} & \textbf{TPR} & \textbf{FPR} & \textbf{Pre} & \textbf{Accuracy}\\ \hline
Cantina \cite{zhang:2007:cantina} & - & - & 2,100 & 19 & English & no learning & 0.89 & 0.03 & 0.212 & 0.969 \\
Cantina+ \cite{xiang:2011:cantina} & 464 & 235 & 1,868 & 940 & several & time & 95.5\% & 1.3\% & 96.4\% & 97\% \\
Whittaker \textit{et al.} \cite{whittaker:2010:large} & 9,284,711 & 103,683 & 1,499,109 & 16,967 & several & time & 91.5\% & 0.01\% & 98.9\% & 99.9\% \\
Thomas \textit{et al.} \cite{thomas:2011:design} & - & - & 500,000 & 50,000 & several & cross-validation & 73.4\% & 0.3\% & 96.1\% & 86.6\% \\
Xiang \textit{et al.} \cite{Xiang:2009:hybrid} & - & - & 7,906 & 3,543 & several & no learning & 90\% & 1.9\% & 95.7\% & 95.5\% \\
Ramesh \textit{et al.} \cite{ramesh:2014:efficious} & - & - & 1,200 & 3,374 & top Alexa & no learning & 99.6\% & 0.5\% & 99.8\% & 99.6\% \\
Ma \textit{et al.} \cite{Ma:2009:beyond} & 15,000 & 20,500 & 15,000 & 20,500 & DMOZ & learning= testing & 92.4\% & 0.1\% & 99.8\% & 95.5\% \\
Chen \textit{et al.} \cite{Chen:2014:anti} & - & - & 404 & 1,945 & top Alexa & cross-validation & 100\% & 0.7\% & 99.2\% & 99.4\% \\ \hline
Our method & 4,531 & 1,036 & 100,000 & 1,216 & English & time & 95.8\% & 0.05\% & 95.6\% & 99.9\% \\
Our method & 4,531 & 1,036 & 150,000 & 1,216 & several & time & 95.8\% & 0.1\% & 85.7\% & 99.8\% \\
Our method & - & - & 4,531 & 1,036 & English & cross-validation & 95.7\% & 0.1\% & 99.1\% & 99\% \\

\end{tabular}
\caption{Performance comparison with state-of-the-art}
\label{tab:compare2}
\end{table*}
}

The obfuscation and mimicry characteristics of phishing webpages have been the basis of several solutions proposed for phishing detection and target identification.

\textbf{Phishing webpage detection:}
Analysis of the content \cite{Xiang:2009:hybrid,zhang:2007:cantina} and code execution (\textit{e.g.} the use of javascript, pop-up windows, etc.) \cite{corbetta:2014:eyes,Mohammad:2014:predicting} of a webpage provides relevant information to identify phishing webpages. Some detection methods rely on URL lexical obfuscation characteristics \cite{Blum:2010:lexical,le:2011:phishdef,marchal:2014:phishstorm} and webpage hosting related features \cite{Feroz:2014:examination,Ma:2009:beyond,Ma:2009:identifying} to render a decision about the legitimacy of a webpage. Other methods \cite{stringhini:2013:shady} study the global interactions of users with a given webpage to infer its maliciousness.
The visual similarity of a phishing webpage with its target was also exploited to detect phishs \cite{Chen:2010:detecting,Chen:2014:anti,Medvet:2008:visual}.
These phishing detection based on visual similarity presuppose that a potential target is known a priori. In contrast, our approach is to discover the target. 

Multi-criteria methods \cite{Ma:2009:identifying,thomas:2011:design,whittaker:2010:large} have been proved the most efficient to detect phishing websites.
These techniques use a combination of webpage features (HTML terms, links, frame, etc.), connection features (HTML header, redirection, etc.) and host based features (DNS, IP, ASN, geolocation, etc.) to infer webpage legitimacy. They are implemented as offline systems checking content pointed by URLs to automatically build blacklists. This process induces a delay of several hours \cite{whittaker:2010:large} that is problematic in the context of phishing detection, since phishing attacks have a median lifetime of a few hours \cite{apwg:2015}. In addition, it is reportedly costly \cite{thomas:2011:design} and use \cite{whittaker:2010:large} some proprietary features (\textit{e.g.} Google PageRank \cite{page:1999:pagerank}) preventing usage on the end-user devices. The identification method uses machine learning techniques fed with hundreds of thousands of features. These features are mostly static and learned from training sets containing data such as IP address, Autonomous System Number (ASN), bag-of-words for different data sources (webpage, URL, etc.).
This limits the generalizability of the approach as it requires large training datasets, numbering hundreds of thousand of webpages \cite{thomas:2011:design,whittaker:2010:large}.
 
Other methods focused, as we do, on the study of terms that compose the data sources of a webpage \cite{Blum:2010:lexical,le:2011:phishdef,Marchal:2012:proactive,marchal:2014:phishstorm}. 
Cantina \cite{xiang:2011:cantina,zhang:2007:cantina} was among the first systems to propose a lexical analysis of terms that compose a webpage. In Cantina \cite{zhang:2007:cantina} key terms are selected  using TF-IDF to provide a unique signature of a webpage. Using this signature in a  search engine, Cantina infers the legitimacy of a webpage. A similar method \cite{Xiang:2009:hybrid}, based on TF-IDF and Google search, checks for inconsistency between a webpage identity and the identity it impersonates to identify phish. The main difference between these methods and ours is language independence since these methods rely on TF-IDF computation to infer their keyterms. Moreover, they rely on centralized infrastructure through search engines query to classify a webpage, while we apply a comparison in term usage only between sources that compose a webpage.

Table \ref{tab:compare} presents comparative performances results of our phishing detection system to the most relevant state-of-the-art systems. It presents the size of the testing sets used to evaluate each system and the provenance of the legitimate set, showing how representative the set is.  For example, using popular websites (such as top Alexa sites)~\cite{Chen:2014:anti,ramesh:2014:efficious} as the legitimate set is not representative. The ratio of training to testing instances indicates the scalability of the method and the ratio of legitimate to phishing instances shows the extent to which the experiments represents a real world distribution ($ \approx 100/1$). We also identify the evaluation method (e.g., cross validation vs. training with old data and testing with new data). Finally, we present several metrics for assessing the classification performance. If data for any of the columns were missing from the original paper describing the system, we estimated them.
For comparison purposes, if several experimental setups were proposed in a paper, we selected the most relevant to assess their practical efficacy using the following ordered criteria:

\begin{enumerate}[topsep=0pt,itemsep=-1ex,partopsep=0pt,parsep=1ex]
 \item learning and testing instances are different,
 \item the ratio of legitimate to phishing in the testing set is representative of real world observations ($ \approx 100/1$),
 \item the learning set is older than the testing set,
 \item the false positive rate (FPR) is minimized.
\end{enumerate}

We can see that among the eight most relevant state-of-the-art techniques, only two \cite{Ma:2009:beyond,whittaker:2010:large} have comparable false positive rates to ours ($\leq 0.001$). A low false positive rate is paramount for a phishing detection technique, since this relates to the proportion of legitimate webpages to which a user will be incorrectly denied.
The technique proposed by Ma \textit{et al.} \cite{Ma:2009:beyond} has a lower accuracy than in our system ($0.955<0.999$). In addition, they use a testing set that does not represent real world distribution (3 legs/4 phishs) and use a cross-validation that does not assess scalability of the approach with a 1/1 ratio for learning to testing instances. 
Whittaker \textit{et al.} \cite{whittaker:2010:large} report results similar to us in several metrics. However, they use a \emph{huge training set (\textgreater 9M instances)} and their test set is actually \emph{smaller} than the training set (a sixth, at 1.5M)! Scalability and language-independence are likely to be poor since they use 100,000 mostly static features (bag-of-words).

In contrast to the state-of-the-art in phishing detection, our solution is language independent, scalable, requires training sets that are much smaller than the test sets and does not rely on real-time access to external sources, while performing better than or as well as the state-of-the-art.


\iftargetid
\textbf{Target identification:}
This is a challenging problem that has not been addressed widely in the literature. One proposal \cite{ramesh:2014:efficious} was to use a similar technique as Cantina with keywords retrieval and Google search to discover a list of potential target as the top results of the search, but the authors do not report accuracy figures for target identification.
\iffullversion
The HREF links of the page are further studied to infer if the webpage is likely to be a phish or not and deduce the target. 
\fi
HREF links have been used to build community graphs of webpages. By counting the mutual links between two webpages and further performing visual similarity analysis between suspicious webpages, Liu \textit{et al.} \cite{liu:2012:anti} identify the target of a given phishing website with an accuracy of 92.1\%. However, this technique is slow because of the need to crawl many additional websites to build the community graph.
Conditional Random Fields and Latent Dirichlet Allocation (LDA) \cite{blei:2003:latent} have been applied to phishing email content to identify their target \cite{ramanathan:2013:phishing} with a success rate of 88.1\%.

The technique we propose, in contrast to previous techniques is language independent for keyterms inference. It is as efficient as any state-of-the-art solutions achieving a maximum success rate of 90.5-97.3\%.
\fi

 

\else

\begin{table*}[tbh]
\centering
\begin{tabular}{l r r c c c c l l l l }
& \multicolumn{2}{c}{\textbf{Testing set}} & \textbf{Legitimate} & \textbf{Train} & \textbf{Leg} & & & & & \\
\textbf{Technique} &  \textbf{Legitimate} & \textbf{Phish} & \textbf{set} & \textbf{/Test} & \textbf{/Phish}  & \textbf{Evaluation}  & \textbf{FPR}  & \textbf{Pre.} & \textbf{Recall} & \textbf{Acc.}\\ \hline
Cantina \cite{zhang:2007:cantina} & 2,100 & 19 & English & - & \textbf{\textcolor{green}{110/1}} & no learning & \textbf{\textcolor{red}{0.03}} & \textbf{\textcolor{red}{0.212}} & 0.89 & 0.969 \\
Cantina+ \cite{xiang:2011:cantina} & 1,868 & 940 & several & 1/4 & 2/1 & old/new & \textbf{\textcolor{red}{0.013}} & 0.964 & 0.955 & 0.97 \\
Xiang \textit{et al.} \cite{Xiang:2009:hybrid} & 7,906 & 3,543 & several & - & 2/1 & no learning & \textbf{\textcolor{red}{0.019}} & 0.957 & 0.9 & 0.955 \\
Ma \textit{et al.} \cite{Ma:2009:beyond} &  15,000 & 20,500 &  DMOZ & 1/1 & \textbf{\textcolor{red}{3/4}} & cross-valid& 0.001 & \textbf{\textcolor{green}{0.998}} & 0.924 & 0.955 \\
Whittaker \textit{et al.} \cite{whittaker:2010:large} &  1,499,109 & 16,967 & several & \textbf{\textcolor{red}{6/1}} & \textbf{\textcolor{green}{90/1}} & old/new & \textbf{\textcolor{green}{0.0001}} & 0.989 & 0.915 & \textbf{\textcolor{green}{0.999}} \\
Thomas \textit{et al.} \cite{thomas:2011:design} & 500,000 & 500,000 & several & 4/1 & 1/1 & cross-valid & 0.003 & 0.961 & \textbf{\textcolor{red}{0.734}} & \textbf{\textcolor{red}{0.866}} \\
Ramesh \textit{et al.} \cite{ramesh:2014:efficious} & 1,200 & 3,374 & \textbf{\textcolor{red}{top Alexa}} & - & \textbf{\textcolor{red}{1/3}} & no learning & \textbf{\textcolor{red}{0.005}} & \textbf{\textcolor{green}{0.998}}  & \textbf{\textcolor{green}{0.996}} & 0.996 \\
Chen \textit{et al.} \cite{Chen:2014:anti} & 404 & 1,945 & \textbf{\textcolor{red}{top Alexa}} & \textbf{\textcolor{red}{9/1}} & \textbf{\textcolor{red}{1/5}} & cross-valid & \textbf{\textcolor{red}{0.007}} & 0.992& \textbf{\textcolor{green}{1}}  & 0.994 \\ \hline
Our method & 100,000 & 1,216 & English & \textbf{\textcolor{green}{1/18}} & \textbf{\textcolor{green}{85/1}} & old/new & \textbf{\textcolor{green}{0.0005}} & 0.956  & 0.958 & \textbf{\textcolor{green}{0.999}} \\
Our method & 150,000 & 1,216 & several & \textbf{\textcolor{green}{1/27}} & \textbf{\textcolor{green}{125/1}} & old/new & 0.001 & 0.857  & 0.958 & \textbf{\textcolor{green}{0.998}} 
\end{tabular}
\caption{Phishing detection system performances comparison}
\label{tab:compare}
\end{table*}

The obfuscation and mimicry characteristics of phishing webpages have been the basis of several solutions proposed for phishing detection and target identification.

\textbf{Phishing webpage detection:}
Analysis of the content \cite{Xiang:2009:hybrid,zhang:2007:cantina} and code execution (\textit{e.g.} the use of javascript, pop-up windows, etc.) \cite{Mohammad:2014:predicting} of a webpage provides relevant information to identify phishing webpages. Some detection methods rely on URL lexical obfuscation characteristics \cite{le:2011:phishdef,marchal:2014:phishstorm} and webpage hosting related features \cite{Feroz:2014:examination,Ma:2009:beyond} to render a decision about the legitimacy of a webpage.
The visual similarity of a phishing webpage with its target was also exploited to detect phishs \cite{Chen:2014:anti,Medvet:2008:visual}.
Phishing detection based on visual similarity presuppose that a potential target is known a priori. In contrast, our approach is to discover the target. 

Multi-criteria methods \cite{thomas:2011:design,whittaker:2010:large} have been proved the most efficient to detect phishing websites.
These techniques use a combination of webpage features (HTML terms, links, frame, etc.), connection features (HTML header, redirection, etc.) and host based features (DNS, IP, ASN, geolocation, etc.) to infer webpage legitimacy. They are implemented as offline systems checking content pointed by URLs to automatically build blacklists. This process induces a delay of several hours \cite{whittaker:2010:large} that is problematic in the context of phishing detection, since phishing attacks have a median lifetime of a few hours \cite{apwg:2015}. In addition, it is reportedly costly \cite{thomas:2011:design} and use \cite{whittaker:2010:large} some proprietary features preventing usage on the end-user devices. The identification method uses machine learning techniques fed with hundreds of thousands of features. These features are mostly static and learned from training sets containing data such as IP address, Autonomous System Number (ASN), bag-of-words for different data sources (webpage, URL, etc.).
This limits the generalizability of the approach as it requires large training datasets, numbering hundreds of thousand of webpages \cite{whittaker:2010:large}.
 
Other methods focused, as we do, on the study of terms that compose the data sources of a webpage \cite{le:2011:phishdef,marchal:2014:phishstorm}. 
Cantina \cite{xiang:2011:cantina,zhang:2007:cantina} was among the first systems to propose a lexical analysis of terms that compose a webpage. In Cantina \cite{zhang:2007:cantina} key terms are selected  using TF-IDF to provide a unique signature of a webpage. Using this signature in a  search engine, Cantina infers the legitimacy of a webpage. A similar method \cite{Xiang:2009:hybrid}, based on TF-IDF and Google search, checks for inconsistency between a webpage identity and the identity it impersonates to identify phish. The main difference between these methods and ours is language independence since these methods rely on TF-IDF computation to infer their keyterms.

Table \ref{tab:compare} presents comparative performances results of our phishing detection system to the most relevant state-of-the-art systems. It presents the size of the testing sets used to evaluate each system and the provenance of the legitimate set, showing how representative the set is. For example, using popular websites (such as top Alexa sites)~\cite{Chen:2014:anti,ramesh:2014:efficious} as the legitimate set is not representative. The ratio of training to testing instances indicates the scalability of the method and the ratio of legitimate to phishing instances shows the extent to which the experiments represents a real world distribution ($ \approx 100/1$) \cite{whittaker:2010:large,zhang:2007:cantina}. We also identify the evaluation method (e.g., cross validation vs. training with old data and testing with new data). Finally, we present several metrics for assessing the classification performance. If data for any of the columns were missing from the original paper describing the system, we estimated them.
For comparison purposes, if several experimental setups were proposed in a paper, we selected the most relevant to assess their practical efficacy using the following ordered criteria:

\begin{enumerate}[topsep=0pt,itemsep=-1ex,partopsep=0pt,parsep=1ex]
 \item learning and testing instances are different,
 \item the ratio of legitimate to phishing in the testing set is representative of real world observations ($ \approx 100/1$),
 \item the learning set is older than the testing set,
 \item the false positive rate (FPR) is minimized.
\end{enumerate}

We can see that among the eight most relevant state-of-the-art techniques, only two \cite{Ma:2009:beyond,whittaker:2010:large} have comparable false positive rates to ours ($\leq 0.001$). A low false positive rate is paramount for a phishing detection technique, since this relates to the proportion of legitimate webpages to which a user will be incorrectly denied.
The technique proposed by Ma \textit{et al.} \cite{Ma:2009:beyond} has a lower accuracy than in our system ($0.955<0.999$). In addition, they use a testing set that does not represent real world distribution (3 legs/4 phishs) and use a cross-validation that does not assess scalability of the approach with a 1/1 ratio for learning to testing instances. 
Whittaker \textit{et al.} \cite{whittaker:2010:large} report results similar to us in several metrics. However, they use a \emph{huge training set (\textgreater 9M instances)} and their test set is actually \emph{smaller} than the training set (a sixth, at 1.5M)! Scalability and language/brand independence are likely to be poor since they use 100,000 mostly static features (bag-of-words).

In contrast to the state-of-the-art in phishing detection, our solution is language independent, scalable, requires much smaller training sets than test sets, and does not rely on real-time access to external sources, while performing better than or as well as the state-of-the-art.


\textbf{Target identification:}
One proposal \cite{ramesh:2014:efficious} was to use a similar technique as Cantina with keywords retrieval and Google search to discover a list of potential target as the top results of the search, but the authors do not report accuracy figures for target identification.
HREF links have been used to build community graphs of webpages. By counting the mutual links between two webpages and further performing visual similarity analysis between suspicious webpages, Liu \textit{et al.} \cite{liu:2012:anti} identify the target of a given phishing website with an accuracy of 92.1\%. However, this technique is slow because of the need to crawl many additional websites to build the community graph.
Conditional Random Fields and Latent Dirichlet Allocation (LDA) \cite{blei:2003:latent} have been applied to phishing email content to identify their target \cite{ramanathan:2013:phishing} with a success rate of 88.1\%.

The technique we propose, in contrast to previous techniques is language independent for keyterms inference. It is as efficient as any state-of-the-art solutions achieving a maximum success rate of 90.5-97.3\%.
\fi

\section{Conclusion}
We presented novel techniques for efficiently and economically
identifying phishing webpages and their targets. By using a set of
features that capture inherent limitations that phishers face, our
system has excellent performance and scalability while requiring much
smaller amounts of training data. We have also implemented a fully client-side phishing prevention 
browser add-on implementing this technique~\cite{armano:2016:real}.

\section{Acknowledgments}
This work was supported in part by the Academy of Finland (grant
274951) and Intel Collaborative Research Center for Secure Computing
(ICRI-SC). We thank Craig Olinsky, Alex Ott and Edward Dixon for
valuable discussions.

\bibliographystyle{IEEEtran}

\end{document}